\documentclass{article}
\usepackage{jheppub,esint,shuffle,psfrag}
 \usepackage[utf8]{inputenc}

\usepackage{varioref}
\usepackage{amsmath,amsfonts,amsthm,mathrsfs}
\usepackage{enumerate}
\usepackage{fancyvrb}
\usepackage{verbatim}
\usepackage{wrapfig}
\usepackage{appendix}
\usepackage{amstext}
\usepackage{amssymb}
\usepackage{graphicx}
\usepackage{color}
\usepackage{varioref}
\usepackage{multirow,graphics}
\usepackage{epstopdf}
\usepackage{bbm}
\usepackage{slashed}
\usepackage{relsize}

\numberwithin{equation}{section}

\usepackage{tikz}
\usetikzlibrary{plotmarks,calc,decorations, decorations.pathmorphing, patterns}
\usetikzlibrary{arrows}
\tikzstyle dynkin node=[very thick,shape=circle,draw,inner sep=0pt,minimum size=5mm]
\tikzstyle dynkin line=[very thick]
\tikzstyle inverse line=[gray,line width=1.46pt,line cap=round, dash pattern=on 0pt off 2\pgflinewidth]
\tikzstyle red phase=[red,decoration={snake,amplitude=0.1mm,segment length=1.6mm},decorate]
\tikzstyle blue phase=[blue,decoration={snake,amplitude=0.1mm,segment length=0.9mm},decorate]
\tikzstyle green phase=[green,decoration={snake,amplitude=0.1mm,segment length=0.9mm},decorate]
\tikzstyle brown phase=[brown,decoration={snake,amplitude=0.1mm,segment length=0.9mm},decorate]
\newcommand{\boundellipse}[3]
{(#1) ellipse (#2 and #3)
}
\usetikzlibrary{decorations.pathmorphing}
\tikzstyle arrow=[thick,rounded corners=18pt,-latex]
\tikzstyle box=[draw,rounded corners,outer sep=4pt]
\tikzstyle B node=[outer sep=0pt]
\tikzstyle Q node=[inner sep=1pt,outer sep=0pt]
\definecolor{purple_nice}{rgb}{0.4,0.2,0.7}
\definecolor{fuel_blue}{RGB}{42,162,185}

\def\<{\langle}
\def\>{\rangle}

\newcommand{\sym}{${\cal N}=4$ SYM}

\newcommand{\p}{\partial}
\newcommand{\tr}{{\text{tr}}}

\makeatletter
\@ifundefined{usebibtex}{} {}
\makeatother

\def\Det{\text{det}~}

\newcommand{\fl}{\hphantom{.}}

\title{
\Large Basso-Dixon Correlators in Two-Dimensional Fishnet CFT    }

\author[a]{Sergei Derkachov}
\author[b,c]{~Vladimir Kazakov}
\author[d]{~Enrico Olivucci}

\affiliation[a]{St. Petersburg Department of the Steklov Mathematical Institute
of Russian Academy of Sciences,
Fontanka 27, 191023 St. Petersburg, Russia}
\affiliation[b]{Laboratoire de Physique Th\'eorique, D\'epartement de Physique de l'ENS, \'Ecole Normale Sup\'erieure, rue Lhomond 75005
Paris, France}
\affiliation[c]{Universit\'e Paris-VI, PSL Research University, Sorbonne Universit\'es, UPMC Univ. Paris 06, CNRS, 75005 Paris, France}
\affiliation[d]{II. Institut f\"ur Theoretische Physik, Universit\"at Hamburg, Luruper Chaussee 149, 22761
Hamburg, Germany}

\abstract{
   We compute explicitly the two-dimensional version of Basso-Dixon type integrals for the planar 4-point correlation functions given by conformal ``fishnet" Feynman graphs. These diagrams are represented by a fragment of a regular square lattice of power-like propagators,   arising in the recently proposed integrable bi-scalar fishnet CFT.  The formula is derived from first principles, using the formalism of separated variables in integrable \(SL(2,\mathbb{C})\) spin chain. It is generalized to anisotropic fishnet, with different powers for propagators in two directions of the lattice.   
}

\usepackage{esint}
\usepackage{breqn}

\def \tr {\mathop{\rm tr}\nolimits}

\def\numberbysection{\@addtoreset{equation}{section}
                     \def\theequation{\thesection.\arabic{equation}}}

\begin{document}

\hskip9cm\preprint{ LPTENS-04/ZMP-HH/18-25}
\maketitle

\flushbottom

\newpage

\section{Introduction}\label{intro}

\begin{figure}[t]
\centerline{\includegraphics[scale=0.9]{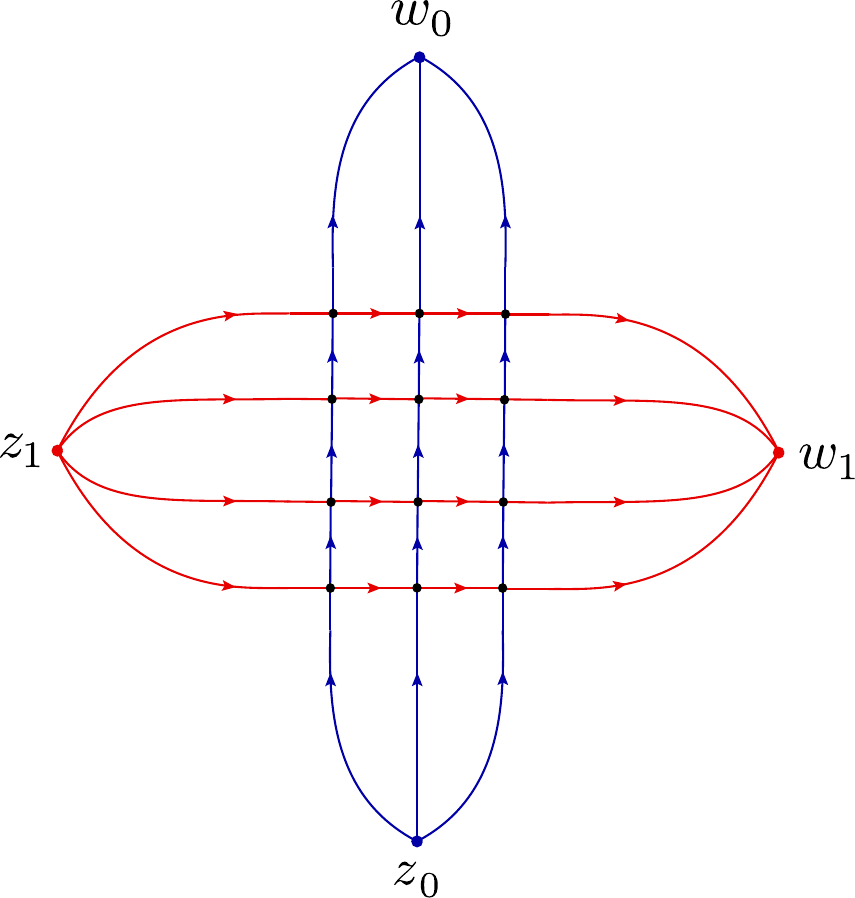}}
\caption{Basso-Dixon type Feynman diagram for \(N=4,\,L=3\).
The propagators have the form \([w-z]^{-\alpha}\) where
  \(\alpha=1/2\pm \omega\) for vertical and horizontal lines, respectively.}
\label{BDgraph}
\end{figure}

Recently, B.~Basso and L.~Dixon obtained an elegant explicit expression for a specific, conformal planar Feynman graph of  fishnet type  
 \cite{Basso:2017jwq}, having \(N\) rows and \(L\) columns, and thus \((N+1)(L+1)-4\) loops. This graph is presented on Fig.\ref{BDgraph}. It has four external fixed coordinates and, similarly to the conformal 4-point functions, has a non-trivial   dependence  on two cross-ratios \(u,v\). This Basso-Dixon (BD) formula
takes the form of an \(N\times N\) determinant of explicitly known ``ladder" integrals~\cite{Usyukina:1993ch,Isaev:2007uy}. It is one of very few examples of explicit results for Feynman graphs with arbitrary many loops.

The BD formula appeared in the context of its application to the four dimensional conformal theory which emerged as a specific double scaling limit of \(\gamma\)-deformed \sym~   theory combining weak coupling and strong imaginary \(\gamma\)-twists~\cite{Gurdogan:2015csr,Caetano:2016ydc}. In particular, in one-coupling reduction of this theory -- the so called bi-scalar, or ``fishnet" CFT -- the BD integral represents indeed a particular single-trace correlation function (described below). In general, the bulk structure of planar graphs in fishnet CFT is that of the regular square lattice of massless propagators. Such a graph represents an integrable two-dimensional statistical mechanical system~\cite{Zamolodchikov1980a} which can be studied via integrable quantum spin chain with the symmetry of 4D conformal group \(SU(2,2)\)~~\cite{Gurdogan:2015csr,Caetano:2016ydc,Gromov:2017cja,Chicherin:2012yn}.

Two of the current authors recently proposed the \(D\)-dimensional generalization of bi-scalar fishnet theory~\cite{Kazakov:2018qez}. Its action is given in terms of two interacting complex \(N_c\times N_c\) matrix scalar fields \(X(x),Z(x)\):
\begin{align}
    \label{bi-scalarL}
\mathcal S= N_c\int d^Dx\,\tr\Big(   X^\dagger (-\partial^\mu\partial_\mu)^{D/4+\omega} X+Z^\dagger (-\partial^\mu\partial_\mu)^{D/4-\omega}Z
+ (4\pi)^2 \xi^2  X^\dagger  Z^\dagger X Z\Big)\,,
  \end{align}
where \(\omega\)  is an  arbitrary ``anisotropy"   parameter producing different powers of propagators along two axis of the fishnet square lattice  and     \(\xi\) is the coupling constant.~\footnote{Strictly speaking, to tune this theory to the conformal point at any \(\xi\) one should add to this action the double-trace interactions with specific \(\xi\)-dependent
 coefficients~\cite{Tseytlin:1999ii,Dymarsky:2005uh,Fokken:2013aea,Fokken:2014soa,Sieg:2016vap,Gromov:2017cja}.} At \(D=4,\, \omega=0\) it reduces to the local bi-scalar action following from the double scaling limit of \sym~\cite{Gurdogan:2015csr}.
The BD-type integral corresponds to the following single-trace correlation function: \begin{align}
I_{L,N}^{{BD}}(z_0,z_1,w_0,w_1) = \left<\tr\left(X^L(z_0)Z^N(z_1) X^{\dagger L}(w_0)Z^{\dagger N}(w_1)\right)\right>.
\label{bi-localO}\end{align}  It is easy to see that, due to the chiral nature of interaction of two scalars, this correlation function is given in the planar limit by a single, fishnet-type planar graph of BD-type drawn in Fig.\ref{BDgraph}. Explicitly, this Feynman graph is given by expression
\begin{align}\fl\label{BDint}
&I_{L,N}^{{BD}}(z_0,z_1,w_0,w_1)=\notag\\
=&\int \,\prod_{(l,n)\in {\cal L}_{L,N}}
d^D z_{l,n} \left(\prod_{(l,n)\in {\cal L}_{L,N+1}} \frac{1}{|z_{l,n-1}-z_{l,n}|^{D/2+2\omega}}\right)\left( \prod_{(l,n)\in {\cal L}_{L+1,N}} \frac{1}{|z_{l-1,n}-z_{l,n}|^{D/2-2\omega}}\right)
\end{align}
%
%
\begin{figure}[t]
\centerline{\includegraphics[scale=1]{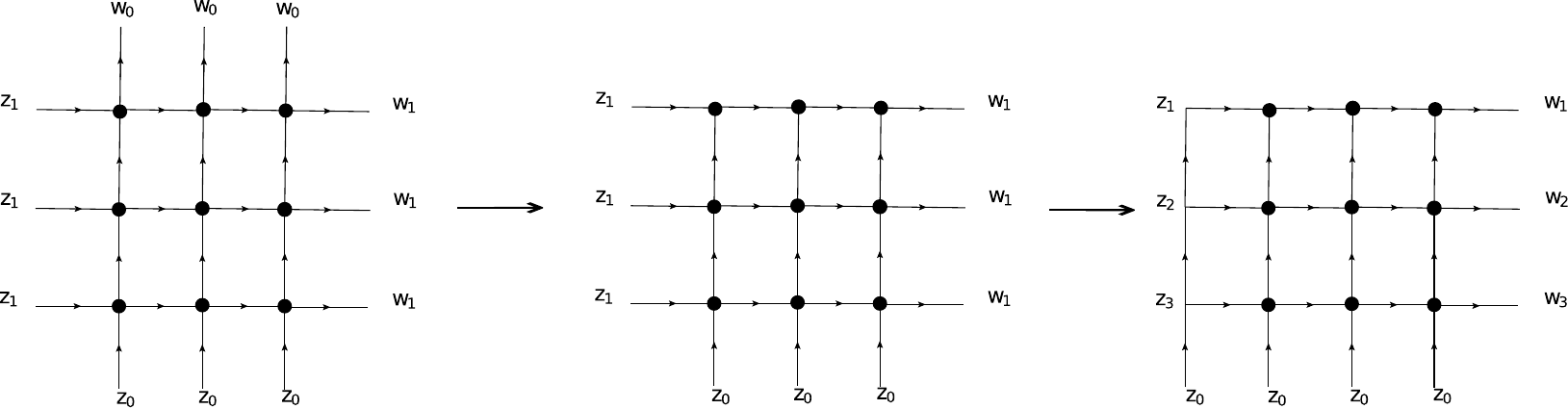}}
\caption{Basso-Dixon type diagram \(I_{L,N}^{{BD}}(z_0,z_1,w_0,w_1)\) (on the left), its reduction  \(G_{L,N}(\textbf{z}|\textbf{w})\) (in the middle) and generalization  \(D_{L,N}(\textbf{z}|\textbf{w})\) (on the right) described in sec.\ref{transfGraph}. We integrate only the coordinates in the  vertices marked by black blobs. Sending \(w_0\to\infty\) in the original Basso-Dixon type diagram, we remove the upper row of propagators and obtain the reduced diagram (in the middle). Using  conformal invariance of the original  graph (on the left), we can always restore it from the graph on the right, by inversion and shift of coordinates \(w_1,z_1,z_0\).  Further on, we generalize the middle diagram by splitting the end point coordinates of left  and right columns  of external propagators,  to separate coordinates \(z_1\to(z_1,z_2,\dots,z_N)\) and \(w_1\to(w_1,w_2,\dots,w_N)\), and then add at the left a column of vertical propagators \([z_i-z_{i+1}]^{-\gamma}\), thus getting the generalized configuration (on the right).    
  }
\label{BDtransf}
\end{figure}
where we have \(N\cdot L\)  integration variables belonging to the \(L\times N\) lattice of positive integers \(\mathcal{L}_{L,N}=\{1\leq L \leq L, 1\leq n \leq N\}\), and we  take equal coordinates at each of the four boundaries of this rectangular lattice: \(\{z_{j,0}=z_0,\,\, z_{j,N+1}=w_0,\,\, z_{0,k}=z_1,\,\, z_{L+1,k}=w_1\}\) are imposed for \(j=1,...,L\) and \(k=1,...,N\). 

This integral was computed explicitly  in \(D=4\),  for ``isotropic" case \(\omega=0\), in    \cite{Basso:2017jwq}. The derivation is based on certain assumptions, typical for the \(S\)-matrix bootstrap methods inherited from the integrability of planar \sym~\cite{Beisert:2010jr}. It would be important to derive this formula from the first principles, based on the conformal spin chain interpretation of fishnet graphs, but in four dimensions such a derivation is so far missing.

    In this paper, we will derive from the first principles the explicit expression for the two-dimensional analogue, \(D=2\), of Basso-Dixon formula for the ``fishnet" Feynman integral\footnote{Here and in the following we adopt the notation \([z-w]^\alpha\equiv (z-w)^\alpha (z^*-w^*)^{\bar\alpha}\) for propagators, see App.\ref{app_A1} for details.}
\begin{eqnarray}\fl\label{BDint2D}
&&I_{L,N}^{{BD}}(z_0,z_1,w_0,w_1)= \\ \,&&=\int\, \prod_{(l,n)\in {\cal L}_{L,N}}
d^D z_{l,n} \left(\prod_{(l,n)\in {\mathcal{L}_{L,N+1}}} \frac{1}{[z_{l,n-1}-z_{l,n}]^{\gamma}}\right)\left( \prod_{(l,n)\in {\mathcal{L} _{L+1,N}}} \frac{1}{[z_{l-1,n}-z_{l,n}]^{1-\gamma}}\right), \nonumber
\end{eqnarray} where the coordinates \((z_0,z_1,w_0,w_1)\)   are defined as after the eq.\eqref{BDint}. We took here propagators transforming in the spinless complementary series of representations  (\(\bar \gamma= \gamma \in (0,1)\)) under \(SL(2,\mathbb{C})\) group action \eqref{bar_s}. The  propagators  for \(D=2,\,\, \omega=\gamma-1/2\) are \( [w-z]^{-1/2 \mp \omega}\), where \(\mp\)
  is chosen for vertical and horizontal lines, i.e.  for the fields \(X,Z\), respectively. 

Our derivation is based on integrable \(SL(2,\mathbb{C}) \) spin chain methods worked out in~\cite{Belitsky:2014rba,Derkachov:2014gya,Derkachov:2001yn}, using the Sklyanin separation  of  variables (SoV) method ~\cite{Sklyanin:1984sb,Sklyanin:1995bm,Sklyanin:1991ss}. The result can be presented in explicit form, in terms of \(N \times N \) determinant of a matrix with the elements which are explicitly computed in terms of  hypergeometric functions of cross-ratios\footnote{Or alternatively, due to the obvious \(L\leftrightarrow N\) symmetry of the integral, in terms of the \((L-1)\times(L-1)\) determinant of the same matrix elements, which will depend only on \(L+N\) combination.}.  Our main formula looks as follows:

\begin{align}\label{BDBfinal}
I^{BD}_{L,N}(z_0,z_1,w_0,w_1)=
 \,\frac{[z_1-z_0]^{(\gamma-1)N}[w_1-w_0]^{(\gamma-1)N}}{[z_0-w_0]^{(\gamma-1)N+\gamma L}} \,[\eta]^{\frac{\gamma-1}{2} N} B^{(\gamma)}_{L,N}\left(\eta \right)
\end{align}
where\begin{align}
B^{(\gamma)}_{L,N}(\eta,\bar\eta)&=
(2\pi)^{-N} \pi^{-N^2} \,\underset{1\le j,k\le N}{\Det} \left[(\eta\p_\eta)^{i-1}(\bar\eta\p_{\bar\eta})^{k-1}I^{(\gamma)}_{N+L}(\eta,\bar\eta)\right],\qquad \eta=\frac{z_0-w_1}{w_1-w_0}\frac{z_1-w_0}{z_0-z_1},\label{BNL}
 \end{align}
and
\begin{align}
I^{(\gamma)}_M(\eta,\bar\eta) =&\frac{2\pi^{M+1}}
{(M-1)!\,[\eta]^{\frac{\gamma-1}{2}}} \frac{\Gamma^M(\gamma)}{\Gamma^M(1-\gamma)}\, \times\notag\\
\times \,& \left.\partial_{\varepsilon}^{M-1}\right|_{\varepsilon=0}\left[
\, \frac{\Gamma^M(1-\gamma-\varepsilon)}{\Gamma^M(\gamma+\varepsilon)} \frac{\Gamma^M(1+\varepsilon)}{\Gamma^M(1-\varepsilon)}\,
[\eta]^{-\varepsilon}\,\, \left|{}_{M+1}F_M \left(\left.\begin{matrix}1-\gamma-\varepsilon &  \cdots &1-\gamma-\varepsilon & \; 1 \\ \;\; 1-\varepsilon & \cdots & 1-\varepsilon &\end{matrix}\; \right| \eta \right)\right|^2\right] \nonumber.
\end{align}
Formula \eqref{BNL} is also generalized  in sections \ref{BDcomputation}, \ref{ladder_sect} to the principal series representations of \(SL(2,\mathbb{C})\), see \eqref{bar_s}.
 
In the next section, we will  define the basic building blocks for construction of the  Basso-Dixon configuration in operatorial way. In section~\ref{Lambda-operator}, we will introduce the generalized ``graph-building" operator related to the transfer-matrix of the integrable open  \(SL(2,C)\) quantum spin chain. We will  diagonalize there  this operator by means of the SoV method and  describe  the full system of  its eigenfunctions. In section~\ref{BDcomputation}    the result for 2d Basso-Dixon-like \(N\times L\) graph will be presented in terms of an \(N\times N\) determinant of the matrix constructed from \(1 \times M\) such graph called the ladder graph. In  section~\ref{ladder_sect}, the ladder graph will be computed explicitly, in terms of the hypergeometric functions and their derivatives, thus completing the explicit result for the full two-dimensional Basso-Dixon-like \(N\times L\) graph  presented above. The ladder graph is employed to compute the so-called simple wheel graph in two dimensions. A particular case of \(N=L=1\) (the two-dimensional ``cross" graph) will be explicitly given in terms of the elliptic functions of the cross ratio.

\section{Transformations of Basso-Dixon type graph and \(L\leftrightarrow N\) duality}\label{transfGraph}
In order to apply powerful methods of \(SL(2,\mathbb{C}) \) spin chain integrability, such as the separation of variables (SoV), we will use the conformal symmetry to reduce the BD graph on Fig.\ref{BDgraph} to a more convenient quantity for our purposes. First of all, we  send \(w_0\to\infty\) and  drop the corresponding propagators  containing this variable:
\begin{align*}&I^{BD}_{L,N}(z_0,z_1,w_0,w_1) 
\underset{w_0\to\infty}{\to} [w_0]^{-\gamma L}G_{L,N}(z_1,w_1|z_0) \\ 
&\text{where}\quad  G_{L,N}(z_1,w_1|z_0)=\\&=\int\,  \prod_{l=1}^L\,\prod_{n=1}^{N}       \, d^2z_{ln}\left( \prod_{\substack{1 \leq l \leq L\\ 1\leq n \leq N}}\, \frac{1}{[z_{l,n-1}-z_{l,n}]^{\gamma}
\times[ z_{l,n}-z_{l+1,n}]^{1-\gamma}}\right)\prod_{n=0}^{N-1}\, \frac{1}{[z_{0,n}-z_{1,n}]^{1-\gamma}},
\end{align*}  where we take \(\{z_{j,0}=z_0,\,\, z_{0,k}=z_1,\,\, z_{L+1,k}=w_1\}\)  for \(j=1,...,L\) and \(k=1,...,N\).
 We can always restore the original quantity  \(I^{BD}_{L,N}(z_0,z_1,w_0,w_1)\) from \(G_{L,N}(z_1,w_1|z_0)\), presented on  Fig.\ref{BDtransf}(middle), using  the conformal symmetry of   \(I^{BD}_{L,N}\)  , i.e. by  applying the inversion+shift transformation and thus getting the original quantity \eqref{BDBfinal} (see Appendix~\ref{reductions} for derivation and  examples). 

Now we are going to generalize the quantity \(G_{L,N}(z_1,w_1|z_0)\), in order to apply the integrability techniques. To that end, we introduce a more general quantity drawn on Fig.\ref{BDtransf}(right):
\begin{align}
\label{GLNzw}
  D_{L,N}(z_0)({\boldsymbol{z}}|\boldsymbol{w}) =\int\,  \prod_{l=1}^L\,\prod_{n=1}^{N}       \, d^2z_{ln}\left( \prod_{(l,n)\in {\cal L}_{L+1,N}}\, \frac{1}{[z_{l-1,n-1}-z_{l-1,n}]^{\gamma}
\times [z_{l-1,n}-z_{l,n}]^{1-\gamma}}\right),
\end{align} 
where all the external legs on the left and on the right of Fig.\ref{BDtransf}(left)  have different coordinates: \(\{z_{j,0}=z_0,\,\, z_{0,k}=z_k,\,\, z_{L+1,k}=w_k\}\)  for \(j=0,1,...,L\) and \(k=1,...,N\).
 We introduced in the r.h.s. of \eqref{GLNzw} the  vector notations: \({\boldsymbol{z}}=\{z_1,z_2,\dots,z_N\}, 
 \boldsymbol{w}=\{w_1,w_2,\dots,w_N\}\). Notice that, after point-splitting, we  multiplied, for the future convenience, the middle diagram of Fig.\ref{BDtransf} by the vertical propagators on the left, without altering the essential part of the quantity, since the coordinates in the left column are exterior and they  are not integrated.

\begin{figure}[t]
\centerline{\includegraphics[scale=1.2]{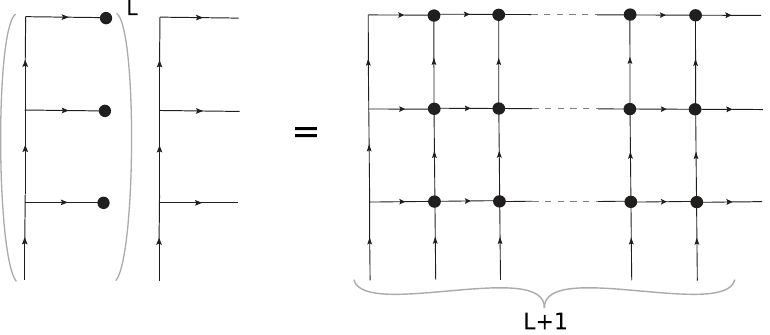}}
\caption{The ``comb" transfer matrix for an open spin chain of length \(N\) (\(N=3\) on the picture) is applied \(L\) times to itself as an integral kernel. The resulting structure is a Fishnet of the type of fig.\ref{BDtransf}(right) with \(L+1\) vertical and \(3\) horizontal lines.}
\label{BDiter}
\end{figure}

The last expression \eqref{GLNzw}, representing the diagram on the right of Fig.\ref{BDtransf}, is the most appropriate for the application of integrability methods. Namely, we can represent it as a consecutive action of a ``comb" transfer matrix ``building" the graph, as shown on the Fig.\ref{BDiter}.
%
In the next section, we will define yet a more general transfer matrix \(\Lambda_N(x)({\boldsymbol{z}}|\boldsymbol{w})   \) depending on a spectral parameter \(x\) and diagonalize it by means   of  eigenfunctions using separation of variable (SoV) method of Sklyanin. The lattice of propagators can be  inhomogeneous in \(L\)-direction, since each transfer matrix, corresponding to an open spin chain of length \(N\)  ``building" the BD configuration  by \(L\) consecutive applications, as on  Fig.\ref{BDiter},
 can have its own spectral parameter.   Its particular, homogeneous case will give the explicit formula for \(2D\)  BD graph.~\footnote{Still containing the anisotropy parameter \(\gamma\).}

Now we will comment on the  obvious \(L\leftrightarrow N\)  duality of the original BD diagram:
\begin{equation}
I^{BD}_{L,N}(z_0,z_1,w_0,w_1;\gamma)=I^{BD}_{N,L}(z_1,w_0,z_0,w_1,1-\gamma),
\end{equation} where we explicitly introduced among the arguments the anisotropy parameter \(\gamma\).  It is useful to represent the same quantity in a more explicitly conformally symmetric way:
\begin{equation}
I^{BD}_{L,N}(z_0,z_1,w_0,w_1;\gamma)=[w_0-z_0]^{-L \gamma}[w_1-z_1]^{N(\gamma-1)}\, [\eta]^{N\frac{\gamma-1}{2}}[1-\eta]^{N(1-\gamma)} B^{(\gamma)}_{L,N}(\eta).   
\label{4pointform}
\end{equation}\,\,  Then the \(L\leftrightarrow N\)  duality reads as follows:\begin{equation}
\label{duality_intro}
 B^{\left(1-\gamma \right)}_{N,L}(1/\eta)\, = \,[\eta]^{\frac{\gamma}{2}(N+L)-\frac{N}{2}}[1-\eta]^{-(N+L)\gamma +N}\,B^{(\gamma)}_{L,N}(\eta).
\end{equation}

\section{``Graph building" operator \(\Lambda_N(x|z_0)\) and its diagonalization}
\label{Lambda-operator}

\begin{figure}[t]
\centerline{\includegraphics[width=0.8\linewidth]{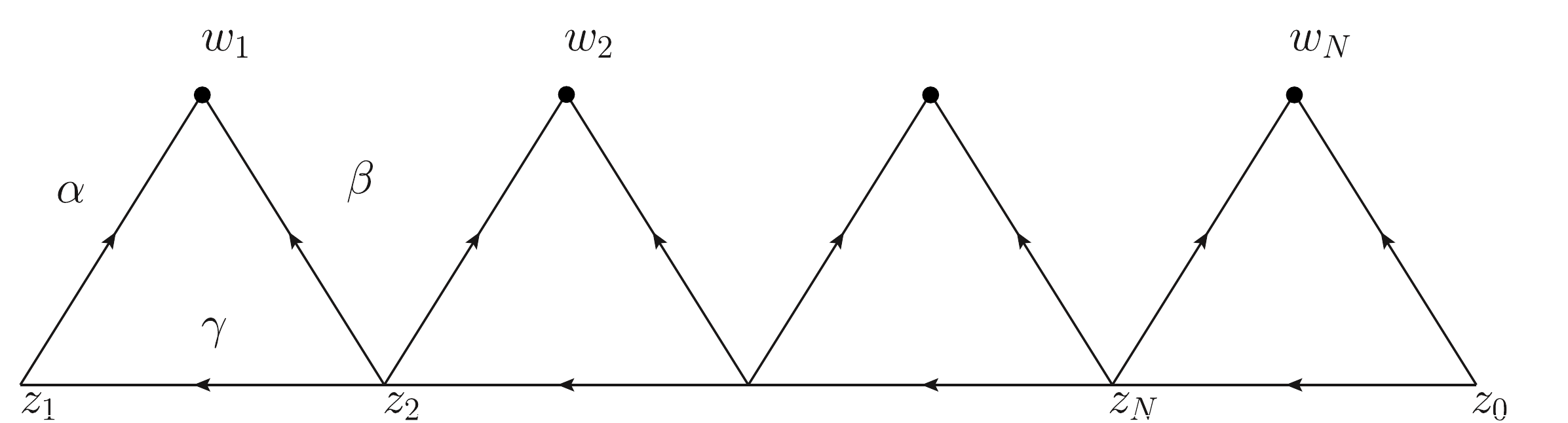}}
\caption{The diagrammatic representation for the kernel of $\Lambda_{N}(y|z_0)$.
The arrow with index $\alpha$ from $z$ to $w$ stands for $[w-z]^{-\alpha}$.
The indices are given by the following expressions: $\alpha=1-s-y$, $\beta=1-s+y$, $\gamma=2s-1$.}
\label{R1N0}
\end{figure}

\begin{figure}[t]
\centerline{\includegraphics[width=1.1\linewidth]{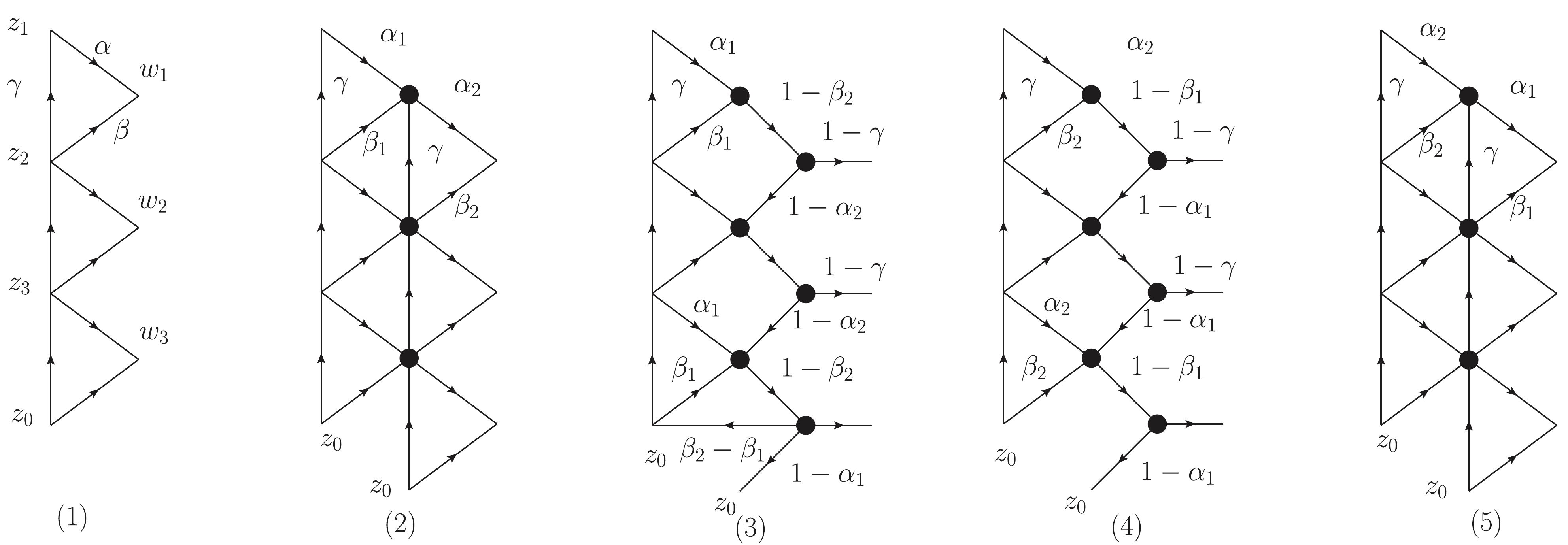}}
\caption{
The proof of  commutation relation  \eqref{com} for two operators $\Lambda_N(y|z_0)$:
(1) The diagram for the kernel of $\Lambda_{3}(y|z_0)$.
(2) The diagram for $\Lambda_{3}(y_1|z_0)\,\Lambda_{3}(y_2|z_0)$: $\alpha_1=1-s-y_1$, $\alpha_2=1-s-y_2$, $\beta_1=1-s+y_1$, $\beta_2=1-s+y_2$, $\gamma=2s-1$.
(3) Triangle-star transformations inside the right column of triangles, leading to  $\Lambda_{3}(y_2)$
(4) Movement of the line with index $\beta_2-\beta_1$ upstairs using cross relations.
(5) Star-triangle transformations back to
$\Lambda_{3}(y_2|z_0)\,\Lambda_{3}(y_1|z_0)$.
}
\label{R1R2}
\end{figure}

Our main goal in the rest of this paper is  the computation of the quantity \(B^{(\gamma)}_{L,N}(\eta)\) directly related to the BD integral by \eqref{4pointform}. To that end, we define a more general transfer matrix of an open \(SL(2,\mathbb{C})\) spin chain, building the generalized BD graph. The explicit computations will be carried out for values of \(\gamma\) corresponding to the principal series of representations of \(SL(2,\mathbb{C})\). Then the original quantity \eqref{BDint2D} is obtained by analytic continuation to real \(\gamma=\frac{1}{2} + \omega\) in the final result.\\ First of all,  we fix  our parameters:
\begin{itemize}
\item
Definition of the conformal spin:
\begin{eqnarray}\label{bar_s}\fl
 s=\frac{1+n_s}2+i\nu_s \ \,,\ \bar s=\frac{1-n_s}2+i\nu_s 
\end{eqnarray}
where \(n_s \in \mathbb{Z} \) is the \(SO(2)\) spin and \(\nu_s \in \mathbb{R}\), so that \(1+ 2 i \nu_s\) is the scaling dimension in the principal series of representations \cite{Tod:1977harm}.
\item
Definition of the $x_k$-parameters which will play the role of spin chain inhomogenieties in spectral parameter, and then also of Sklyanin  separated variables:
\begin{eqnarray}\label{bar_x}\fl
x_k=\frac{n_k}2+i\nu_k\ \,,\ \bar{x}_k =-\frac{n_k}2+i\nu_k
\end{eqnarray}
where \(n_k \in \mathbb{Z} \) and \(\nu_k \in \mathbb{R}\).
\item The spin $s$ and the parameter $x$ (or \(y\)) will enter
almost everywhere in  special combinations~\footnote{In what follows, we will always use the notation \(y\),\(y_k\) when the separated variables appear as spectral parameters of an operator, while \(x\),\(x_k\) when they label an eigenfunction. Both notations refer to objects of the kind \eqref{bar_x}.}, so that
for simplicity we shall use the shorthand notations and define
the $\alpha\,,\beta\,,\gamma$-parameters
\begin{eqnarray}\fl
\alpha=1-s-y\ \,,\ \beta = 1-s+y\ \,,\ \gamma = 2s-1\\
\bar\alpha=1-\bar{s}-\bar{y}\ \,,\
\bar\beta = 1-\bar s+\bar y\ \,,\ \bar\gamma = 2\bar s-1
\end{eqnarray}
\end{itemize}
Now let us define the integral operator $\Lambda_N(y|z_0)$ by its explicit  action  on a function
$\Phi(z_1\,,\ldots,z_N)$  by the formula
\begin{eqnarray}
\label{gen_comb}
\left[\Lambda_N(y|z_0)\Phi\right](z_1\,,\ldots,z_N\,,z_0) &=&
\,\prod_{k=1}^{N}
[z_k-z_{k+1}]^{-\gamma}\times  \\
&\times&\int d^2 w_1\cdots d^2 w_N\,
 \fl\nonumber
\prod_{k=1}^{N}[w_k-z_k]^{-\alpha}[w_k-z_{k+1}]^{-\beta}
\,\Phi(w_1\,,\ldots\,,w_N,z_0)\, ,
\end{eqnarray}
where by definition $z_{N+1} = z_0$, and we introduced the symbol \([z]^\alpha\equiv z^\alpha (z^*)^{\bar\alpha}\) (see the details for this notation in App.~\ref{app_A1}).
Note that the operator $\Lambda_{N}(y|z_0)$ maps the function of  $N$ variables
 to the function of $N+1$ variables, but the last variable $z_0$ plays some special role of an external variable.
The diagrammatic representation for the kernel
of the integral operator $\Lambda_N(y|z_0)$ is shown schematically on the Fig.\ref{R1N0}.
%
The operators $\Lambda_N(y|z_0)$ form a commutative family and the
proof of the commutation relation
\begin{eqnarray}\label{com}
\Lambda_N(y_1|z_0)\,\Lambda_N(y_2|z_0) =
\Lambda_N(y_2|z_0)\,\Lambda_N(y_1|z_0)\,
\end{eqnarray}
is equivalent to the proof of the corresponding relation for the kernels which is demonstrated on the Fig.\ref{R1R2}. The proof is presented there diagrammatically, with the help of cross relation~(\ref{Cross}).
 In this way, we proved the integrability of our open spin chain since both operators on each side of the last relation contain different spectral parameter, \(y_1\) or \(y_2\).   %

We shall use the similar notations $\Lambda_{k}(y)$ for $k=2\,,\ldots\,,N-1$ for operators whose action on a function $\Phi(z_1\,,\ldots\,,z_k)$ is defined in a similar way
\begin{eqnarray}
\label{Lambdaop}\fl
\left[\Lambda_{k}(y)\Phi\right](z_1\,,\ldots,z_k,z_{k+1})& =&\,\prod_{i=1}^{k}
[z_i-z_{i+1}]^{-\gamma}\,\times
\\ \fl\nonumber
&\times &\int d^2 w_1\cdots d^2 w_{k}\,\prod_{i=1}^{k}[w_i-z_i]^{-\alpha}[w_i-z_{i+1}]^{-\beta}\,
\,\Phi(w_1\,,\ldots\,,w_k)\,,
\end{eqnarray}
The variable \(z_{k+1}\) plays here a special role and
the diagrammatic representation for the
kernel of $\Lambda_{k}(y)$ is the same as for $\Lambda_{N}(y|z_0)$ with the evident substitutions $N\to k$
and $z_0\to z_{k+1}$.

\subsection{Eigenfunctions of the operator $\Lambda_N(y|z_0)$}

The eigenfunctions of the operator $\Lambda_N(y|z_0)$ are constructed
explicitly and they admit the following representation
\begin{eqnarray}\fl\label{norm}
\Psi({\boldsymbol{x}}|\boldsymbol{z})
= \tilde{\Lambda}_{N-1}\left(x_1\right)\,
\tilde{\Lambda}_{N-2}\left(x_2\right)\cdots \tilde{\Lambda}_{1}\left(x_{N-1}\right)\,[z_{1}-z_{0}]^{-s+x_N}
\end{eqnarray}
where the operators $\tilde{\Lambda}_{N-k}\left(x_k\right)$ differ
from the operators $\Lambda_{N-k}\left(x_k\right)$ by  a simple factor
\begin{eqnarray}
\label{tildeLambda}\fl
\tilde{\Lambda}_{N-k}\left(x_k\right) =
[z_{N-k}-z_{0}]^{-s+x_k}\,r_{N-k}(x_k,\bar x_k)\,\Lambda_{N-k}\left(x_k\right)\, ,
\end{eqnarray}
with \(r_{N-k}\) defined according to
\begin{align}
\label{rfactor}
r_k(x,\bar x)=& \left(\frac{\Gamma(1-\bar s+ \bar x)\Gamma(1-s+x)}{\Gamma(s+x)\Gamma(\bar s - \bar x)}\right)^{k-1}\, .
\end{align}
and where we introduce a shorthand vector notation for the whole set of variables
\begin{eqnarray}\label{XZ}
{\boldsymbol{x}}=\{\boldsymbol{x}_1,
\ldots, \boldsymbol{x}_N
\}
, &\hskip 5mm &\boldsymbol{x}_k=\left(x_k=\frac{n_k}2+i\nu_k\ \,,\ \bar{x}_k =-\frac{n_k}2+i\nu_k\right)
\nonumber\\
\boldsymbol{z}=\{{z}_1,
\ldots,{z}_N\}, &\hskip 5mm & z_k\,\in\, \mathbb{C}\end{eqnarray}
The presence of the pre-factor \eqref{rfactor} in the definition of \(\tilde{\Lambda}_{N-k}(x)\) operators \eqref{tildeLambda} is crucial to prove the exchange relation
\begin{eqnarray}\label{com}
\tilde{\Lambda}_n(x_1)\,\tilde{\Lambda}_{n-1}(x_2) =
\tilde{\Lambda}_{n}(x_2) \,\tilde{\Lambda}_{n-1}(x_1)\, ,
\end{eqnarray} from which follows that \(\Psi({\boldsymbol{x}}|\boldsymbol{z})\) are symmetric functions of the \(\boldsymbol{x}\)-variables 
\begin{align}
\Psi({\boldsymbol{x}}|\boldsymbol{z})\,=\,\Psi(x_1,\dots\,x_k,\dots x_h,\dots, x_N|\boldsymbol{z})\,=\,\Psi(x_1,\dots\,x_h,\dots x_k,\dots, x_N|\boldsymbol{z})\, .
\end{align}\\
The vector of variables $\boldsymbol{x}$ is used as quantum numbers (separated variables) to label  the eigenfunction
and $\boldsymbol{z}$ is the set of complex coordinates in our initial representation.
We will prove that
\begin{eqnarray}\label{eigen}
\Lambda_N(y|z_0)\,\Psi({\boldsymbol{x}}|\boldsymbol{z}) =
\lambda(y,x_1)\cdots\lambda(y,x_N)\,\Psi({\boldsymbol{x}}|\boldsymbol{z})\,,
\end{eqnarray}
where
\begin{eqnarray}\label{lambdaa}
\lambda(y,x_k) = \pi\, a(1-s-y,s+x_k,1+y-x_k)\,(-1)^{[y+x_k]}\,.
\end{eqnarray}
and the function \(a(\alpha,\beta,\gamma)\) is defined in App.\,\ref{app_A1}. We should note that functions $\Psi(\boldsymbol{x}|\boldsymbol{z})$ are generalized eigenfunctions of the operator $A + z_0 B$  where $A,B$ are standard matrix elements of the monodromy matrix \cite{Sklyanin:1991ss, Faddeev:1996iy}.
  
Note that the detailed notation for the eigenfunction should be $\Psi_N({\boldsymbol{x}}|\boldsymbol{z})$ but we shall skip
$N$ almost everywhere for sake of brevity.

\begin{figure}[t]
\centerline{\includegraphics[width=1.0\linewidth]{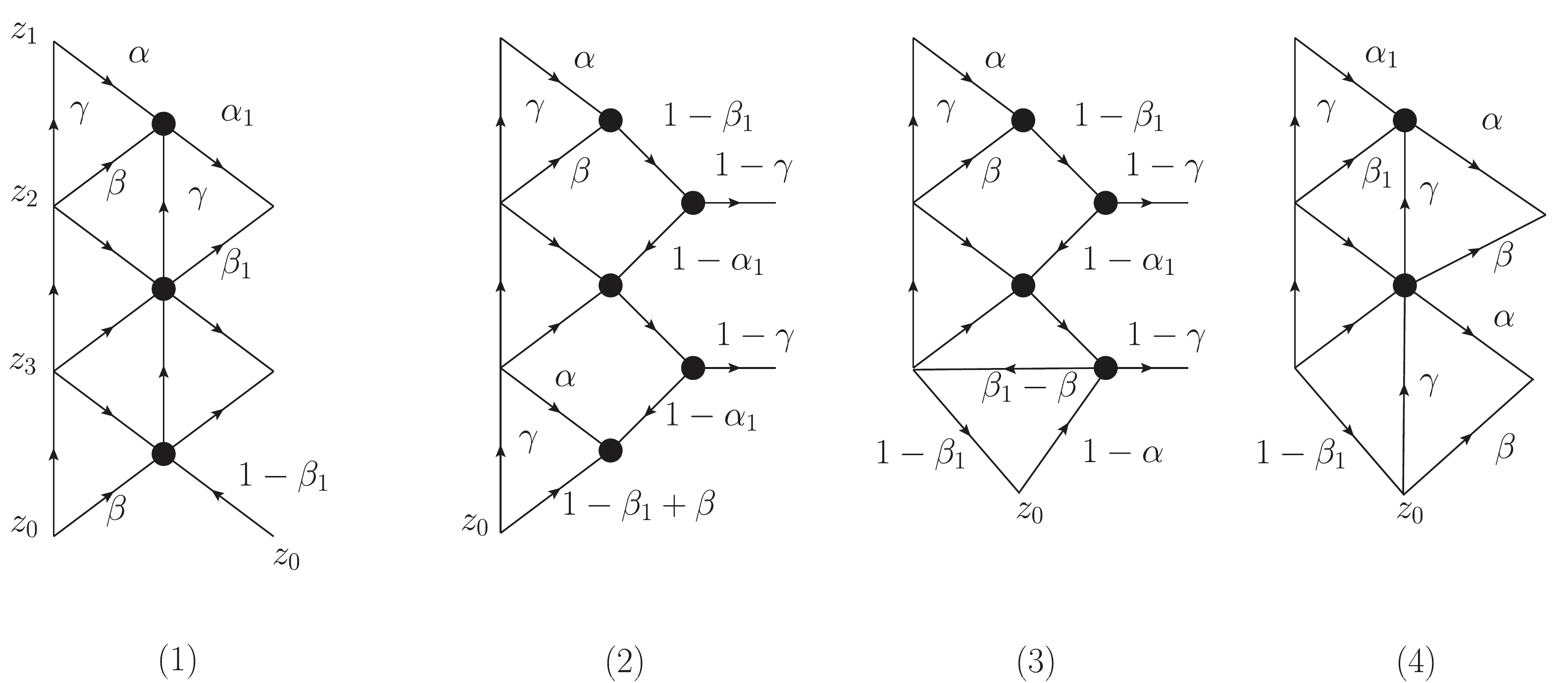}}
\caption{
The proof of diagonalization procedure for the operator \(\Lambda_N(y|z_0)\) for \(N=3\), pushing the operator through the first row of the eigenfunction:    (1) The diagram for $\Lambda_3(y|z_0)\,\tilde{\Lambda}_2(x_1)$: $\alpha=1-s-y$, $\alpha_1=1-s-x_1$, $\beta=1-s+y$, $\beta_1=1-s+x_1$, $\gamma=2s-1$.
(2) Star-triangle transformations inside $\tilde{\Lambda}_2(x_1)$ and two
lines $\beta$ and $1-\beta_1$ ending at $z_0$ joint to the one line
(3) Movement of the line with index $\beta_1-\beta$ upstairs using cross relations leads to
$\tilde{\Lambda}_2(x_1)\,\Lambda_2(y|z_0)$, (4).
}
\label{RR}
\end{figure}

In the simplest case $N=1$ we have
\begin{eqnarray}\fl\nonumber
\Psi(x_1|z_1) &=& [z_{1}-z_{0}]^{-s+x_1}\,,
\\ \fl
\label{N1}
\Lambda_1(y|z_0)\,[z_{1}-z_{0}]^{-s+x_1} &=&
\lambda(y,x_{1})[z_{1}-z_{0}]^{-s+x_1}\,.
\end{eqnarray}
The relation~(\ref{N1}) can be derived by using the chain integration rule~(\ref{Chain}).
The general proof of the relations~\eqref{eigen}-\eqref{lambdaa} is based on the exchange relation
\begin{eqnarray}\fl
\label{exchQA}
\Lambda_N(y|z_0)\, \tilde{\Lambda}_{N-1}(x_1) =
\lambda(y,x_1)\,
\tilde{\Lambda}_{N-1}(x_1)\,
\Lambda_{N-1}(y|z_0)
\end{eqnarray}
The proof of the relation~(\ref{exchQA}) for $N=3$
is shown in Fig.\ref{RR} and the generalization is obvious.
Notice that after exchange, the operator defining the eigenfunction enters with the reduced length \(N\) of the effective spin chain. 
Using the exchange relation step by step it is easy to derive the
formula
\begin{eqnarray}\nonumber
\Lambda_N(y|z_0) \tilde{\Lambda}_{N-1}\left(x_1\right)\,
\tilde{\Lambda}_{N-2}\left(x_2\right)\cdots \tilde{\Lambda}_{1}\left(x_{N-1}\right)= \\
\lambda(y,x_1)\,\lambda(y,x_2)\cdots\lambda(y,x_{N-1})\,
\tilde{\Lambda}_{N-1}\left(x_1\right)\,
\tilde{\Lambda}_{N-2}\left(x_2\right)\cdots \tilde{\Lambda}_{1}\left(x_{N-1}\right)\,\Lambda_1(y|z_0)\,.
\end{eqnarray}
Then the proof that $\Psi({\boldsymbol{x}}|\boldsymbol{z})$ from~(\ref{norm}) is eigenfunction of the operator $\Lambda_N(x|z_0)$ with the eigenvalues given by \eqref{eigen}  is reduced
to the relation~(\ref{N1}) in the form\footnote{This computation, based on uniqueness relation, can also be checked at \(n_k=0,1\) conwith the software \cite{Preti:2018vog}.}
\begin{eqnarray}\nonumber
\Lambda_1(y|z_0)\,[z_{1}-z_{0}]^{-s+x_N} =
\lambda(y,x_{N})[z_{1}-z_{0}]^{-s+x_N}\,.
\end{eqnarray}

We will see that these  eigenfunctions form the complete  orthonormal basis. Using them, as well as   the explicit eigenvalues of \(\Lambda_N(y|z_0)\,\) give above, we will  compute the Basso-Dixon type two-dimensional integral.

\subsection{Orthogonality and completeness}

The functions $\Psi({\boldsymbol{x}}|\boldsymbol{z})$ form a
complete orthonormal basis in the Hilbert space $\mathbb{H}_N$.
Any function $\Phi\in \mathbb{H}_N$ can be expanded w.r.t. this basis as follows
\begin{equation}\fl
\Phi(\boldsymbol{z})=\int \mathcal{D}_N\boldsymbol{x}\, \boldsymbol{ \mu}(\boldsymbol{x})\,
C(\boldsymbol{x})\,\Psi({\boldsymbol{x}}|\boldsymbol{z}) \,.
\end{equation}
The symbol $\mathcal{D}_N\boldsymbol{x}$ stands for the measure in the principal series representation of \(SL(2,\mathbb{C})\) group%
\begin{equation}
\mathcal{D}_N\boldsymbol{x}=\prod_{k=1}^N \left(\sum_{n_k=-\infty}^{\infty}\int_{-\infty}^{\infty} d\nu_k\right)\,.
\end{equation}
Depending on the value of  spin in the quantum space, $n_s=s-\bar s$,
the sum over $n_k$ goes over all integers (integer $n_s$)  or half-integers (half-integer $n_s$).  The coefficient function $C(\boldsymbol{x})$
is given by the scalar product
\begin{eqnarray}
C(\boldsymbol{x})= \int\mathrm{d}^{2N}\boldsymbol{z}\,
\overline{\Psi({\boldsymbol{x}}|\boldsymbol{z})}\,
\Phi(\boldsymbol{z})\,.
\end{eqnarray}
The weight function
$\boldsymbol{ \mu}(\boldsymbol{x})$  \begin{eqnarray}
\boldsymbol{\mu}(\boldsymbol{x})=
\frac{(2\pi)^{-N} \pi^{-N^2}}{N!} \prod_{k<j}[x_k-x_j]\,
\end{eqnarray} is the so-called
Sklyanin measure~\cite{Sklyanin:1984sb,Sklyanin:1995bm}.
It is related to the scalar product of the eigenfunctions
\begin{eqnarray}\label{SCS}
\int\mathrm{d}^{2N}\boldsymbol{z}\,
\overline{\Psi({\boldsymbol{x}'}|\boldsymbol{z})}\,
\Psi({\boldsymbol{x}}|\boldsymbol{z}) =
\boldsymbol{\mu}^{-1}(\boldsymbol{x})\,
\delta_N(\boldsymbol{x}-\boldsymbol{x}')\,.
\end{eqnarray}
Here the delta function $\delta_N(\boldsymbol{x}-\boldsymbol{x}')$
is defined as follows:
\begin{eqnarray}
\delta_N(\boldsymbol{x}-\boldsymbol{x}')=
\frac{1}{N!}\sum_{s\in S_N}\delta(\boldsymbol{x}_1-\boldsymbol{x}'_{s(1)})\ldots
\delta(\boldsymbol{x}_N-\boldsymbol{x}'_{s(N)})\,,
\end{eqnarray}
where summation goes over all permutations of $N$ elements and we define
\begin{equation}
\delta(\boldsymbol{x}-\boldsymbol{x}')\equiv
\delta_{n n'} \delta(\nu-\nu')\,.
\end{equation}

These formulae were obtained in~\cite{Derkachov:2001yn,Derkachov:2014gya} and
the corresponding diagrammatic calculations are discussed at
length in these papers.
The completeness condition for the functions
$\Psi({\boldsymbol{x}}|\boldsymbol{z})$ has
the following form
\begin{eqnarray}\fl\label{complet}
\frac{(2\pi)^{-N} \pi^{-N^2}}{N!}\,
\int \mathcal{D}_N \boldsymbol{x}\,
\prod_{k<j}[x_k-x_j]\,\Psi({\boldsymbol{x}}|\boldsymbol{z})\,
\overline{\Psi({\boldsymbol{x}}|\boldsymbol{z}')} = \prod_{k=1}^N\delta^2(\vec{z}_k-\vec{z}'_k)\,.
\end{eqnarray} A similar formula was proven in the case of \(SL(2,\mathbb{R})\) Toda spin chain by~\cite{Kozlowski:2014jka}, in the case of modular XXZ magnet in \cite{Derkachov:2018lyz} and for \(b\)-Whittaker functions in \cite{shapiro}. It is commonly believed to work for our \(SL(2,\mathbb{C})\) spin chain as well, though the proof is still missing.

\section{SoV representation of generalized Basso-Dixon diagrams and reductions}

\label{BDcomputation}

We have now the necessary instrumentary to reduce the Basso-Dixon type Feynman integrals to the SoV form. First we present the most general, inhomogeneous generalization of our construction and then reduce it to homogeneous anisotropic, or even isotropic case. The last one will be the \(2D\) analogue of the standard fishnet graph considered in \(D=4\) dimensions  in~\cite{Basso:2017jwq}.
 We will suggest for it an explicit determinant representation.
\subsection{SoV representation for general inhomogeneous lattice}
\begin{figure}[t]
\centerline{\includegraphics[width=0.7\linewidth]{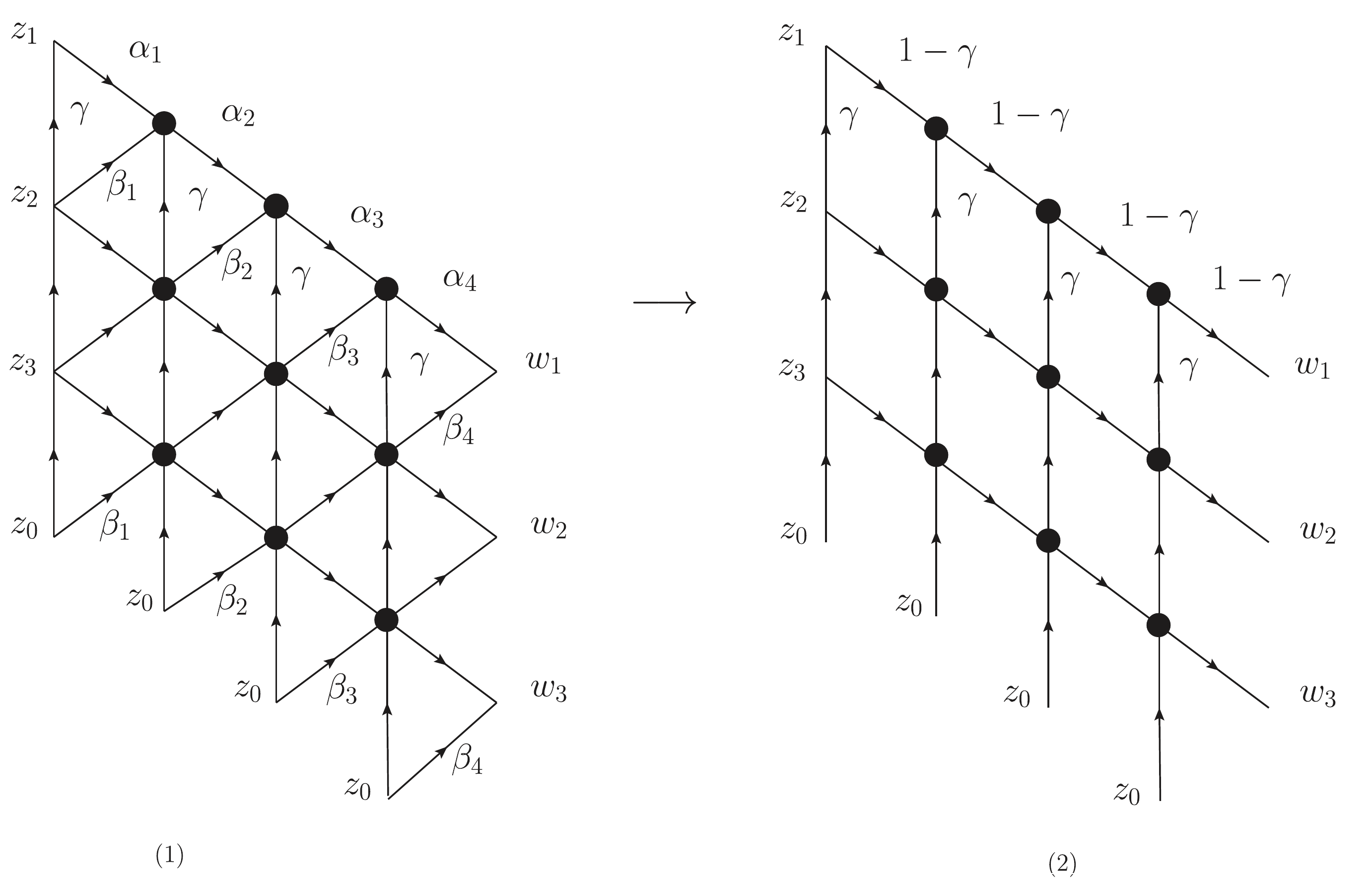}}
\caption{
(1) The diagram for \(\Lambda_3(y_1|z_0)\Lambda_3(y_2|z_0)
\Lambda_3(y_3|z_0)\Lambda_3(y_4|z_0)\): \(\alpha_k=1-s-y_k\), \(\beta_k=1-s+y_k\), \(\gamma=2s-1\).
(2) Reduction of the diagram for \(y_k \to s-1\), or  \(\beta_k \to 0\).}
\label{diagr}
\end{figure}

Using the completeness~(\ref{complet}) and the relation~(\ref{eigen})
we can represent the most general ``graph-generating" kernel,   operator
 \begin{equation}
 \label{Boperat}
 \hat B(y_1,y_2,\cdots,y_L,y_{L+1};z_0)=\Lambda_N(y_1|z_0)\,\Lambda_N(y_2|z_0)
 \cdots\Lambda_N(y_{L+1}|z_0),\end{equation}
which ``builds" a lattice formed by a repeated action of the operator \eqref{gen_comb}.
The integral kernel of the operator \eqref{Boperat} in coordinate representation looks as follows\medskip
\begin{align}\fl\label{1}
& \hat B(y_1,y_2,\cdots,y_L,y_{L+1};z_0)({\boldsymbol{z}}|\boldsymbol{w}) = \nonumber\\
&=\frac{(2\pi)^{-N} \pi^{-N^2}}{N!}
\int \mathcal{D}_N \boldsymbol{x}\,
\prod_{k<j}[x_k-x_j]\,\prod_{k=1}^{N}\prod_{l=1}^{L+1}\lambda(y_l,x_k)\,
\Psi({\boldsymbol{x}}|\boldsymbol{z})\,
\overline{\Psi({\boldsymbol{x}}|\boldsymbol{w})}
\end{align}
 The graphical representation for the left hand side  ~(\ref{1})
for this general case is given in the left picture on Fig.\ref{diagr}.
 This operator is represented there in the form of a lattice with inhomogeneities defined by spectral parameters  \(y_1,y_2,\dots, y_{L+1}\).  Later in this section we will perform the reduction of this formula to the homogeneous lattice of propagators as in the Basso-Dixon integral \eqref{BDint} by taking equal spectral parameters in each column:   \(y_1=y_2=\dots=y_{L+1}= y\), or even a more particular case of homogeneous but anisotropic lattice of propagators (different powers in two directions), putting \(y=s-1\).  But so far we consider the most general configuration.

The diagram in Fig.\ref{diagr} (right) can be reduced to a generalized Basso-Dixon diagram.
First, we have to perform amputation of the most left vertical lines, then
the reduction of all $z_k\to z_1$ in the function
$\Psi({\boldsymbol{x}}|\boldsymbol{z})$ and finally the reduction of all $w_k\to w_1$
in the function
$\overline{\Psi({\boldsymbol{x}}|\boldsymbol{w})}$
in the right hand side of~\eqref{1}.
We will see that such a reduction leads to a significant simplification of the eq.~\eqref{1}, allowing to perform at the end all the integrations and summations over separated variables explicitly.

\begin{figure}[t]
\centerline{\includegraphics[width=1.0\linewidth]{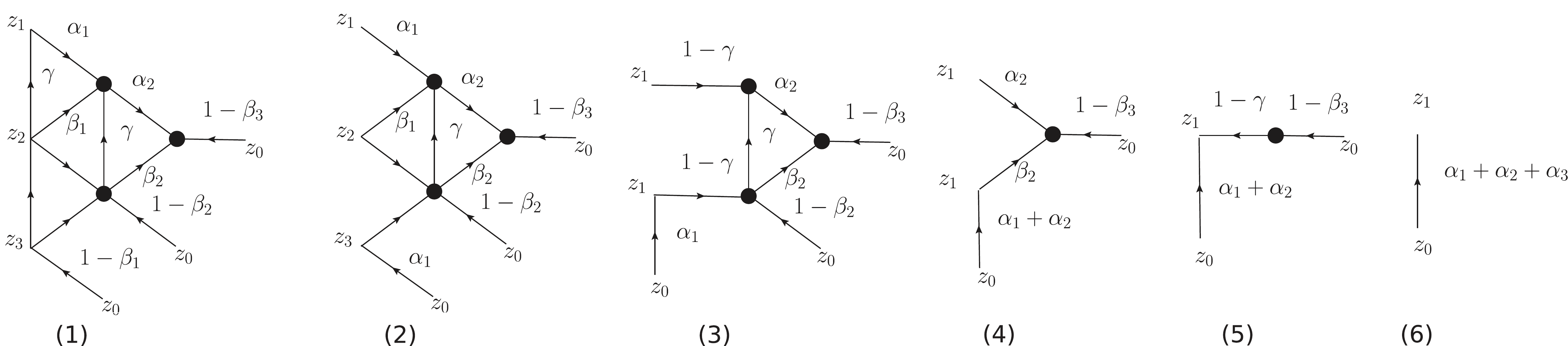}}
\caption{
Amputation of propagators from the eigenfunction $\Psi(x_1,x_2,x_3|z_1,z_2,z_3)$
and then reduction in the limit $z_k\to z_1$ to the simple power $[z_0-z_1]^{-\alpha_1-\alpha_2-\alpha_3}$. We perform amputation of \([z_1-z_2]\) and \([z_2-z_3]\) lines in (1), then (2) we reduce the first row \(z_2,z_3\to z_1\) leading to (3). We can open the triangle in (3) to a star, so that integrations in upper-left, and then lower-left vertex are performed using chain relation and star-triangle relation.  At the next step (4) we join propagators with coinciding coordinates on the left, and performing the last integration (5) via chain relation, the eigenfunction is reduced to a simple line (6).}
\label{BDamp}
\end{figure}

Let us start from the function
$\Psi_N(x_1\,,x_2\ldots x_N|\boldsymbol{z})$.
All the needed steps are illustrated in the Fig.~\ref{BDamp} for $N=3$. Before the reduction $z_k \to z_1$ we have to perform the amputation of the factors
$$
[z_0-z_1]^{-\gamma}\,[z_1-z_2]^{-\gamma}\cdots[z_{N-1}-z_N]^{-\gamma}\,.
$$
After amputation and reduction $z_k \to z_1$ we obtain
the diagram for the action of the operator
$\Lambda_N(x)$ for $x = s-1$ on the function
$\Psi^{(N-1)}(x_2\,,x_3\ldots x_N|\boldsymbol{z})$. It is an eigenfunction for this operator, with the eigenvalue \(\lambda(y_1,x_2)\,\lambda(y_1,x_3)\cdots\lambda(y_1,x_{N})\,
\).
The next step is similar but for a reduced chain $N\to N-1$ and we obtain the
next eigenvalue which is
\(\lambda(y_2,x_3)\,\lambda(y_2,x_4)\cdots\lambda(y_2,x_{N})\,
\), etc.

After all these manipulations we obtain the following formula for the reduction
of the amputated eigenfunction
\begin{eqnarray}\fl\label{NN1}
\prod_{k=0}^{N-1}[z_k-z_{k+1}]^{\gamma}\,
\Psi({\boldsymbol{x}}|\boldsymbol{z}) \to&[z_0-z_1]^{-\alpha_1-\ldots-\alpha_N} \prod_{k=1}^{N}\,
r_{N-k+1}(x_k,\bar x_k)\,\lambda(x_k)^{k-1}\,,
\end{eqnarray}
where we introduced
\begin{align}
\label{lambda}
\lambda(x_k) = \pi a(2-2s,s+x_k,s-x_k)\,(-1)^{[s+x_k]}\, ,
\end{align}
and used the factor \(r_n(x_k, \bar x_k)\) defined in \eqref{rfactor}.

\begin{figure}[t]
\centerline{\includegraphics[width=1.0\linewidth]{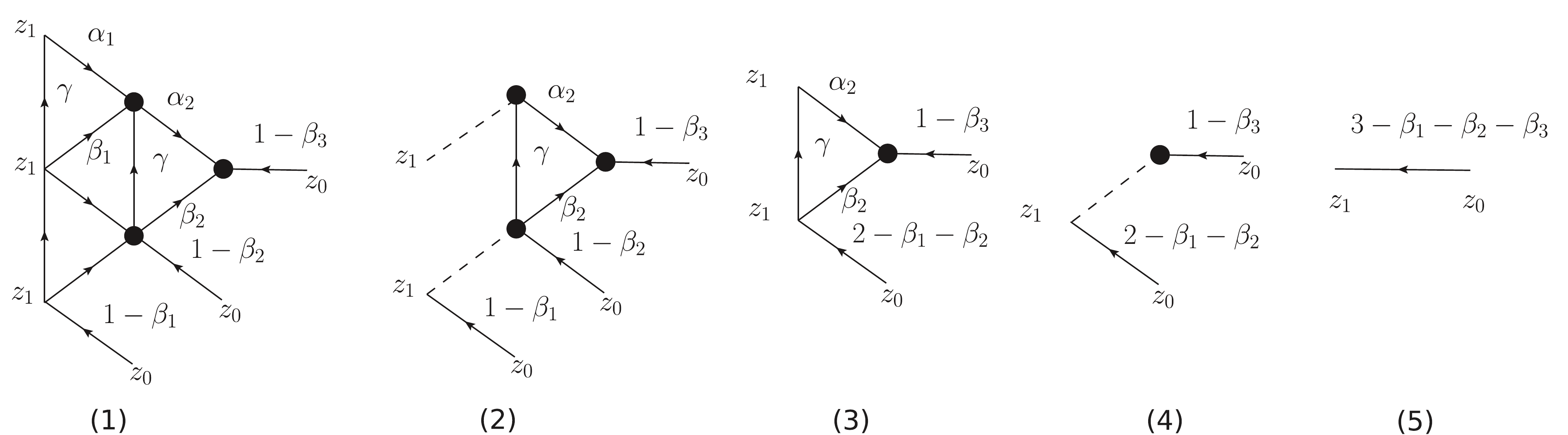}}
\caption{
Reduction of the eigenfunction $\Psi(x_1,x_2,x_3|z_1,z_2,z_3)$ in the limit $z_k\to z_1$ to simple power $[z_0-z_1]^{\beta_1+\beta_2+\beta_3-3}$. Dashed lines stand for \(\delta^{(2)}(z)\), see also \eqref{Delta2}. We reduce \(z_3,z_2\, \to \,z_1\) in (1). By applying triangle-star relations to the first row of triangles (1) we obtain \(\delta\) function kernels. We integrate out \(\delta\) functions (2) and we open the triangle in (3) to a a star and put together the points \(z_1\) obtaining (4). The \(\delta\) function is integrated (4), leading to the full reduction of the eigenfunction to a simple line (5).}
\label{diagrBD}
\end{figure}
The reduction $z_k\to z_1$ for the eigenfunction  $\Psi({\boldsymbol{x}}|\boldsymbol{z})$ without amputations
of the lines is shown step by step
in the Fig.\ref{diagrBD}. First of all we use the star-triangle
relation and reduce the triangle to the corresponding delta-function.
This elementary reduction
\begin{eqnarray}\fl\nonumber
[z_2-z_1]^{-\gamma}\,[w-z_1]^{-\alpha}\,[w-z_2]^{-\beta} \to -\frac{\pi^2}{\gamma\bar{\gamma}}\frac{1}{\lambda(x)}\,\delta^2(z_1-w)
\end{eqnarray}
is shown
on the right in Fig.\ref{BDtransf}.
Using this elementary reduction it is possible to reduce the first layer of the diagram for the general eigenfunction
$\Psi({\boldsymbol{x}}|\boldsymbol{z})$
to the product of the corresponding delta-functions and  $[z_0-z_1]^{\beta_1-1}$ with the coefficient $\left(-\frac{\pi^2}{\gamma\bar{\gamma}}\frac{1}{\lambda(x_1)}\right)^{N-1}$.
After integrations in the corresponding vertices in the second layer all delta-functions disappear so that it is possible to repeat the same procedure.
After all iterations one obtains the following expression for the reduced eigenfunction
\begin{eqnarray}\fl\nonumber
\Psi({\boldsymbol{x}}|\boldsymbol{z}) \to
\prod_{k=1}^N \left(r_{N-k+1}(x_k)\, \left(-\frac{\pi^2}{\gamma\bar{\gamma}}\frac{1}{\lambda(x_k)}\right)^{N-k}\right)
\,[z_0-z_1]^{\beta_1+\ldots+\beta_N-N}\, .
\end{eqnarray}
Note that we have to perform such reduction also in the function $\overline{\Psi({\boldsymbol{x}}|\boldsymbol{w})}$ so that
it remains to perform the complex conjugation and evident substitution
$\boldsymbol{z}\to \boldsymbol{w}$.
Using the rules of the complex conjugation
\begin{eqnarray}\fl
s^{*} = 1-\bar s\,, (x_k)^{*} = -\bar{x}_k \ \ ;\ \ \alpha^{*} = 1-\bar\alpha\,,\beta^{*} = 1-\bar\beta
\,,\gamma^{*} = -\bar\gamma\\ r_k(x_h)^* = r_k(x_h)^{-1}\,\,; \;
\left([z]^{\beta}\right)^{*} = [z]^{1-\beta} \ \ ;\ \, ,
\lambda^{*}(x) = -\frac{\pi^2}{\gamma\bar{\gamma}}\frac{1}{\lambda(x)}
\end{eqnarray}
and substituting $\boldsymbol{z}\to \boldsymbol{w}$ we obtain
\begin{eqnarray}\fl\label{N2}
\overline{\Psi(\boldsymbol{x}|\boldsymbol{w})} \to
\prod_{k=1}^N \left(\lambda^{N-k}(x_k) \,/\,r_{N-k+1}(x_k)\right)\,[z_0-w_1]^{-\beta_1-\ldots-\beta_N}\, .
\end{eqnarray}
Finally, as a result of amputation-reduction on \(\Psi(\boldsymbol{x}|\boldsymbol{z})\) and reduction of \(\overline{\Psi(\boldsymbol{x}|\boldsymbol{w})}\), by the use of~(\ref{NN1}) and~(\ref{N2}) the projector \(\Psi(\boldsymbol{x}|\boldsymbol{z})\overline{\Psi(\boldsymbol{x}|\boldsymbol{w})}\) is transformed into
\begin{eqnarray}
\label{fullred}
\prod_{k=1}^N\lambda^{N-1}(x_k)
\,[z_0-z_1]^{-\alpha_1-\ldots-\alpha_N}
\,[z_0-w_1]^{-\beta_1-\ldots-\beta_N}\,.
\end{eqnarray}
We point out that the way we reduce the \(N\) coordinates \(\boldsymbol{z}=\{z_k\}\) to  a single point in the functions \(\Psi(\boldsymbol{x}|\boldsymbol{z})\) and \(\overline{\Psi(\boldsymbol{x}|\boldsymbol{z})}\) can be alternatively obtained by inserting the complete basis \eqref{complet} between two \(\Lambda\)-kernels in \eqref{Boperat}, and repeating their diagonalization after the reduction of the last kernel \(\Lambda_N(y_{L+1}|z_0)\) and the amputation and reduction of the first \(\Lambda_N(y_1|z_0)\).\\
From formula \eqref{fullred} we obtain the following representation for
the two-dimensional analogue of generalized  Basso-Dixon diagram:
\begin{eqnarray}\fl\label{BDSOVgen1}
G_{N,L}^{\boldsymbol{y}}(z_1\,,w_1\,,z_0) &=&
\frac{(2\pi)^{-N} \pi^{-N^2}}{N!}\,\int \mathcal{D}_N \boldsymbol{x}\,
\prod_{k<j}[x_k-x_j]\times\,\\
\fl\nonumber
\,&\times&\prod_{k=1}^{N}\,\left(\lambda^{N-1}(x_k)
\,\,\prod_{l=1}^{L+1}\lambda(y_l,x_k)\,
\right)[z_0-z_1]^{-\alpha_1-\ldots-\alpha_N}
\,[z_0-w_1]^{-\beta_1-\ldots-\beta_N}\,.
\end{eqnarray}    We recall that \(\alpha_k=1-s-x_k\,,\,\,\beta_k=1-s+x_k\) and \(x_k=\frac{n_k}2+i\nu_k\ \,,\ \bar{x}_k =-\frac{n_k}2+i\nu_k\).

Introducing the amputated cross ratio\begin{eqnarray}\fl\label{cross-ratio}
\left. \eta \right|_{w_0 \rightarrow \infty} = \,\frac{z_0-w_1}{z_0-z_1}
\end{eqnarray}  we rewrite the last expression for inhomogeneous and anisotropic 2D Basso-Dixon type integral in a concise form \begin{eqnarray}\fl\label{BDSOVgen}
G_{L,N}^{\boldsymbol{y}}(z_1\,,w_1\,,z_0) =\,\left([z_0-z_1]
\,[z_0-w_1]\right)^{N(s-1)}\,B^{\boldsymbol{y}}_{L,N}(\eta)
\end{eqnarray} where \begin{eqnarray}\fl\label{BDSeta}
\,B^{\boldsymbol{y}}_{L,N}(\eta)= \frac{(2\pi)^{-N} \pi^{-N^2}}{N!} \,\int \mathcal{D}_N \boldsymbol{x}\,
\,\prod_{k=1}^{N}\,\left([\eta]^{-x_k}\lambda^{N-1}(x_k)
\,\,\prod_{l=1}^{L+1}\lambda(y_l,x_k)\,
\right)\,\prod_{k<j}[x_k-x_j]\,.
\end{eqnarray}and by superscript \(\boldsymbol{y}\) we mean the vector of inhomogeneity parameters \(\boldsymbol{y}=(y_1,y_2,\dots,y_N)\).

\subsection{Determinant representation}
We notice that in \eqref{BDSeta} we deal with the multiple integral of a special type
which can be transformed, similarly to the eigenvalue reduction of the hermitian one-matrix integral~\cite{Brezin:1977sv,Itzykson:1979fi},   to the determinant form
\begin{equation}\label{detrep}
B^{\boldsymbol{y}}_{L,N}(\eta)=\frac{(2\pi)^{-N} \pi^{-N^2}}{N!} \,\int \mathcal{D}_N \boldsymbol{x}\,
\prod_{k<j}[x_k-x_j]\,\prod_{k=1}^N\,f_{\{y\}}(x_k) =
N!\,\Det M
\end{equation}
where we introduced the momenta
\begin{equation}\label{Mik}
\quad M_{ik} = \int \mathcal{D}x\,\,x^{i-1}\bar{x}^{j-1}f_{\{y\}}(x)
\ ;\ i,k = 1\,,\ldots\,,N
\end{equation}
with the weight function given in our case by the expression
\begin{eqnarray}\fl\label{f}
f_{\{y\}}(x) = [\eta]^{-x}\lambda^{N-1}(x)
\,\,\prod_{l=1}^{L+1}\lambda(y_l,x)=\eta^{-x}\bar \eta^{-\bar x}\lambda^{N-1}(x)
\,\,\prod_{l=1}^{L+1}\lambda(y_l,x)
\end{eqnarray} where \(\lambda(x) \) and \(\lambda(y,x)\) are defined in eqs.\eqref{lambda},\eqref{lambdaa}.
 So for any pair of integers \(L,N\) the problem is reduced to the computation of momenta \eqref{Mik}, which we will do explicitly in the section \ref{ladder_sect}  after the  reduction to Basso-Dixon configuration of the general formula \eqref{BDSOVgen}.

\subsection{Reductions}

In particular case, leading to  the  homogenous Basso-Dixon lattice  configuration, we put   \(y_1=y_2=\dots=y_L=y\) and obtain for the reduced quantity
\begin{eqnarray}\fl\label{defLambdaN}
B^{\boldsymbol{y}}(z_0)({\boldsymbol{z}}|\boldsymbol{w})\left|_{y_1=y_2=\dots=y_L=y}\right.\equiv B(y;z_0)({\boldsymbol{z}}|\boldsymbol{w}) =\Lambda^L(y|z_0)
({\boldsymbol{z}}|\boldsymbol{w})
\end{eqnarray}
the following SoV representation:
\begin{eqnarray}\fl\label{genLambda}
B(y;z_0)({\boldsymbol{z}}|\boldsymbol{w}) = \frac{(2\pi)^{-N} \pi^{-N^2}}{N!}
\int \mathcal{D}_N \boldsymbol{x}\,
\prod_{k<j}[x_k-x_j]\,\prod_{k=1}^{N}\lambda^{L}(y,x_k)\,
\Psi({\boldsymbol{x}}|\boldsymbol{z})\,
\overline{\Psi({\boldsymbol{x}}|\boldsymbol{w})}\,.
\end{eqnarray}

The further reduction of this expression, \(\beta_k \to 0\), or \(y_k =y\to s-1\), will lead to anisotropic Basso-Dixon type \(D=2\)  integral \eqref{BDint2D} with parameters \(\gamma=2s-1,\,\bar \gamma=2\bar s -1\). After this reduction we obtain the second diagram in Fig.\ref{diagr}, with the different propagators \([z-z']^{1-2s}\) and \([z-z']^{2s-2}\) in vertical and horizontal directions of the lattice.
In this case, we have to substitute into the  formula~(\ref{1}) representing  this diagram the  reduced eigenvalues
\begin{eqnarray}\nonumber
\lambda(y,x_k) &=& \pi a(1-s-y,s+x_k,1+y-x_k)\,(-1)^{[y+x_k]}\quad\overset{ y = s-1}{\longrightarrow} \\
\longrightarrow\lambda(x_k) &=& \pi a(2-2s,s+x_k,s-x_k)\,(-1)^{[s+x_k]}\,.
\label{lambdaOFx}\end{eqnarray}This leads, after the identification of external coordinates: \(z_k\to z_1,\quad w_k\to w_1\), described above, to the  following representation for
the two-dimensional analog of  (anisotropic) Basso-Dixon diagram \(B_{L\,,N}(\eta)\) in terms of the multiple integral over \(N\) separated variables
\begin{eqnarray}\fl\label{BDSOV}
B_{L,N}(\eta) =
\,\frac{(2\pi)^{-N} \pi^{-N^2}}{N!} \int \mathcal{D}_N \boldsymbol{x}\,
\,\prod_{k=1}^{N}\,[\eta]^{-x_k}\lambda^{N+L}(x_k)\prod_{k<j}[x_k-x_j]\,.\,
\,
\end{eqnarray}

Notice that the parameters of the representation \((s,\overline{s})\) can be chosen in the principal series \eqref{bar_s}, or even in the imaginary strip \(\nu^{(s)}\,\in\,\left(-i/2\, ,0 \right)\) by analytic continuation. With the choice of parameters \(n_s=0\) and \(\nu^{(s)} = -i/4 \pm i\,\omega/2 \) in \eqref{bar_s} we describe the 2D Basso-Dixon type integral with real propagators \(|z-z'|^{-1\mp \omega}\), where \(\pm\) signs corresponds to two different  axis of the square lattice shaped Feynman graph, according to the bi-scalar Lagrangian \eqref{bi-scalarL}. The isotropy of the lattice is restored at \(s=\bar{s}=3/4\), that is \(\omega=0\).

The determinant formula  \eqref{detrep} reads for this reduction as follows \begin{equation}\label{detreduced}
 B^{(\gamma,\bar \gamma )}_{L,N}(\eta)=
(2\pi)^{-N} \pi^{-N^2}\,\underset{1\le j,k\le N}{\Det} m_{jk}\, ,
\end{equation}  where \begin{equation}\label{Mikred}
\quad m_{ik} = \int \mathcal{D}x\,\,x^{i-1}\bar{x}^{j-1}f(x)
\ ;\ i,k = 1\,,\ldots\,,N
\end{equation}
and
\begin{eqnarray}\fl\label{ff}
f(x)= [\eta]^{-x}\lambda^{N+L}(x)=\eta^{-x}\bar \eta^{-\bar x}\lambda^{N+L}(x)
\end{eqnarray} where \(\lambda(x) \) is defined in eqs.\eqref{lambda}.

\section{Explicit computation of ladder integral}
\label{ladder_sect}

In this section, we will explicitly compute the momenta
  \(m_{ik}\) given by \eqref{Mikred}
in terms of hypergeometric functions, which leads to  explicit expressions of Basso-Dixon type integrals via the determinant representation \eqref{detreduced}.
Some details of the derivation can be found in Appendix~\ref{app_B}.

Noticing that \begin{equation}\label{difff}
\quad m_{ik} =(\eta\p_\eta)^{i-1}(\bar\eta\p_{\bar\eta})^{k-1}I_{N+L}\,\,,\quad\text{where}\quad I_{M}=\int\mathcal{D}x \,\eta^{-x}\bar \eta^{-\bar x}\lambda^{M}(x)
\end{equation} we are led to compute the following sum and integral\footnote{We use here and in the following the notation \((-1)^{[\alpha]}\), see \eqref{signnotation}.}:
\begin{eqnarray}\nonumber
&&I_{M}=\int\mathcal{D}x \,\eta^{-x}\bar \eta^{-\bar x}\lambda^{M}(x) =
\pi^M a^M(2-2s)(-1)^{M[s]}\,\int\mathcal{D}x
\,a^M(s+x,s-x)\,(-1)^{M[x]}\eta^{-x}\bar \eta^{-\bar x} =
\\
&&=\pi^M a^M(2-2s)(-1)^{M[s]}\,\sum_{n\in Z}\int_{-\infty}^{+\infty} d \nu\,
\frac{\Gamma^M(1-\bar{s}-\frac{n}{2}+i\nu)\Gamma^M(1-\bar{s}+\frac{n}{2}-i\nu)}
{\Gamma^M(s-\frac{n}{2}-i\nu)\Gamma^M(s+\frac{n}{2}+i\nu)} (-1)^{Mn}\eta^{-\frac{n}{2}-i\nu}{\bar \eta}^{\frac{n}{2}-i\nu}\, , \notag\\
\label{intladder}
\end{eqnarray}
where in the last line we substituted explicit parameters.
\begin{figure}[t]
\centerline{\includegraphics[scale=0.8]{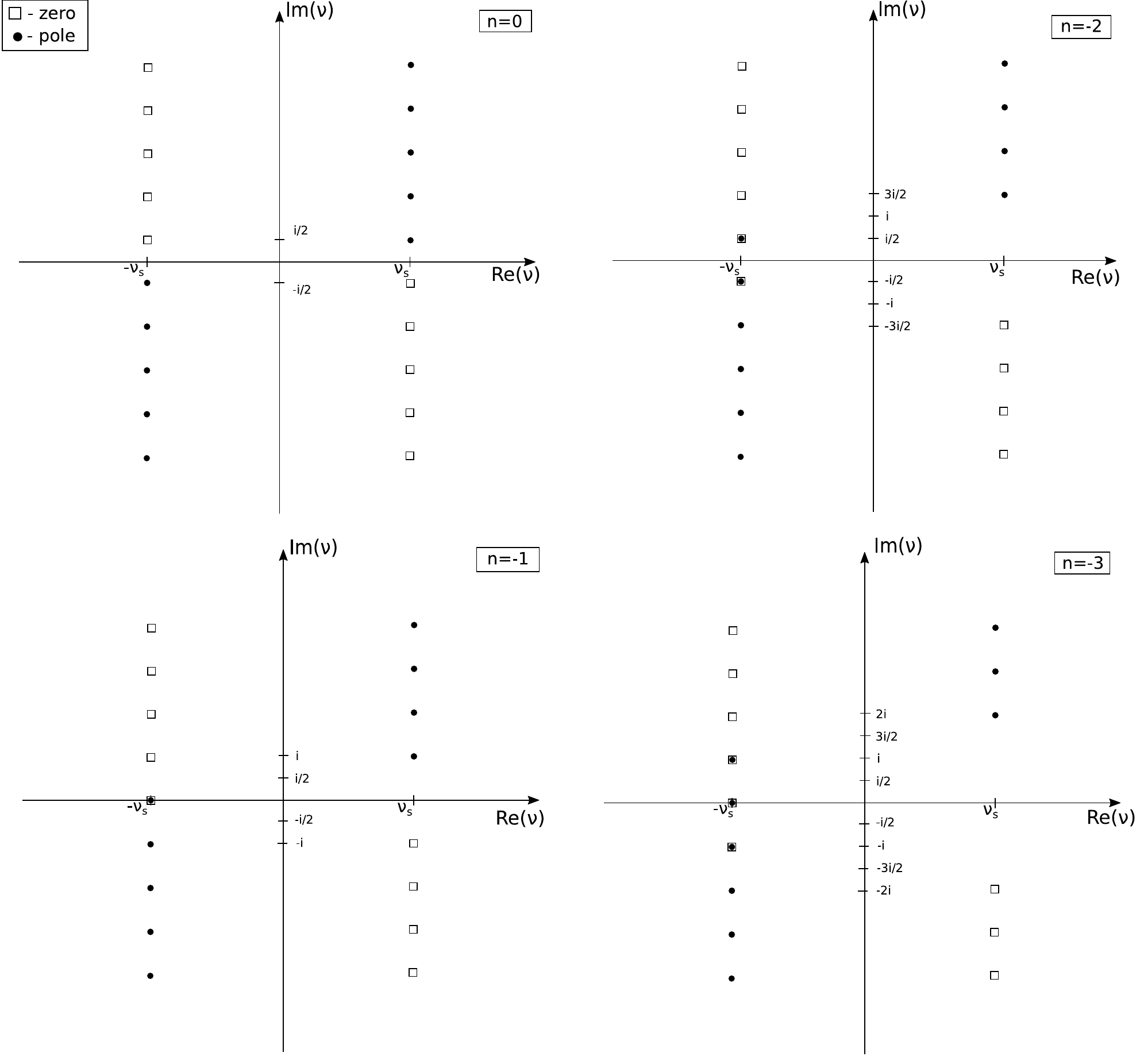}}
\caption{Structure of poles and zeroes of the integrand in \eqref{intladder}, at different values of the discrete variable \(n\), for \(n_s=0\). Superposition of zeroes and poles  occurs in such a way that there is only one semi-infinite series of poles (and zeroes) in  upper- and lower- half-planes.}
\label{poles}
\end{figure}
We will compute the integral over \(\nu\) by residues. The structure of the poles and zeroes is shown in the Fig.~\ref{poles}.
We can close the integration contour on the upper/lower half-plane under the condition \(|\eta|<1\), respectively \(|\eta|>1\), ensuring the exponential suppression of the integrand at \(\pm i \infty \). Consider first the case  \(|\eta|<1\). In the upper half-plane there is one infinite sequence of poles of the order M. After the change of variables $n\to -n+n_s+1$ in the sum over
$n$ and $\nu \to \nu +\nu_s$ in the integral over $\nu$, the integral \eqref{intladder} reads
\begin{eqnarray}\nonumber
&&I_{M}=\frac{\pi^M a^M(2-2s)(-1)^{{M}}}{\eta^s \bar\eta^{\bar s-1} }\,
\sum_{n\in Z}\int_{-\infty}^{+\infty} d \nu\,
\frac{\Gamma^M(2-2\bar{s}-\frac{n}{2}-i\nu)\Gamma^M(\frac{n}{2}+i\nu)}
{\Gamma^M(2s-\frac{n}{2}+i\nu)\Gamma^M(\frac{n}{2}-i\nu)} (-1)^{Mn}\eta^{\frac{n}{2}-i\nu}{\bar \eta}^{-\frac{n}{2}-i\nu}\notag
\label{intladder1}
\end{eqnarray}
We close the contour in the upper half-plane and
calculate the $\nu$-integral as the sum of residues. Due to the mechanism illustrated in fig.\ref{poles}, this is equivalent to take residues at the points $\nu = \frac{in}{2} +ik\,, k=0\,,1\,,2\,,\ldots$, i.e. the series of the poles created by the function $\Gamma^M(\frac{n}{2}+i\nu).$
The residue at the point $\nu = \frac{in}{2} +ik$ can be represented in the following form
\begin{eqnarray*}\label{res}
\mathrm{Res}_{\nu = \frac{in}{2} +ik} = -\frac{i}{(M-1)!}
\left.\partial_{\varepsilon}^{M-1}\right|_{\varepsilon=0}\left[
\frac{\Gamma^M(1+\varepsilon)\Gamma^M(1-\varepsilon)}
{\Gamma^M(2s+\varepsilon)\Gamma^M(1-2s-\varepsilon)}\,
[\eta]^{-\varepsilon}\times \right.
\\
\nonumber
\left.\times\frac{\Gamma^M(1-2s+n+k-\varepsilon)}
{\Gamma^M(n+k-\varepsilon)}
\frac{\Gamma^M(2-2\bar{s}+k-\varepsilon)}
{\Gamma^M(1+k-\varepsilon)} \,\eta^{n+k}\,\bar{\eta}^{k}\right]\,.
\end{eqnarray*}
Using this formula one obtains the following relation
\begin{eqnarray*}
\sum_{n\in Z}\int_{-\infty}^{+\infty} d \nu\,
\frac{\Gamma^M(2-2\bar{s}-\frac{n}{2}-i\nu)\Gamma^M(\frac{n}{2}+i\nu)}
{\Gamma^M(2s-\frac{n}{2}+i\nu)\Gamma^M(\frac{n}{2}-i\nu)} (-1)^{Mn}\eta^{\frac{n}{2}-i\nu}\bar{\eta}^{-\frac{n}{2}-i\nu} = \\
=\frac{2\pi}{(M-1)!}
\left.\partial_{\varepsilon}^{M-1}\right|_{\varepsilon=0}\left[
\frac{\Gamma^M(1+\varepsilon)\Gamma^M(1-\varepsilon)}
{\Gamma^M(2s+\varepsilon)\Gamma^M(1-2s-\varepsilon)}\,
[\eta]^{-\varepsilon}\right.\times\\
\times\left.\sum_{n\in Z}\sum_{k=0}^{+\infty}
\frac{\Gamma^M(1-2s+n+k-\varepsilon)}
{\Gamma^M(n+k-\varepsilon)}
\frac{\Gamma^M(2-2\bar{s}+k-\varepsilon)}
{\Gamma^M(1+k-\varepsilon)} \,\eta^{n+k}\,\bar{\eta}^{k}\right]
\end{eqnarray*}
Remarkably enough, since we take derivative at \(\varepsilon=0\) the last double sum can be equivalently rewritten in a factorized form, setting \(p=n+k-1\)
\begin{eqnarray*}
\eta\,\sum_{p=0}^{+\infty}
\frac{\Gamma^M(2-2s+p-\varepsilon)}
{\Gamma^M(1+p-\varepsilon)}
\,\eta^{p}\,\,
\sum_{k=0}^{+\infty}
\frac{\Gamma^M(2-2\bar{s}+k-\varepsilon)}
{\Gamma^M(1+k-\varepsilon)}\,\bar{\eta}^{k}
\end{eqnarray*}
and we obtain the following expression for the ladder integral
\begin{eqnarray}
\int\mathcal{D}x \lambda^M(x)[\eta]^{-x} &=&
\frac{2\pi^{M+1} a^M(1-\gamma)(-1)^{M}}
{(M-1)!\,[\eta]^{\frac{\gamma-1}{2}}}\,\times\\
\nonumber
&\times&\left.\partial_{\varepsilon}^{M-1}\right|_{\varepsilon=0}
\frac{\Gamma^M(1+\varepsilon)\Gamma^M(1-\varepsilon)}
{\Gamma^M(\gamma+1+\varepsilon)\Gamma^M(-\gamma-\varepsilon)}\,
[\eta]^{-\varepsilon}\,
F_M(1-\gamma \,,\varepsilon|\eta)\,F_M(1-\bar \gamma \,,\varepsilon|\bar{\eta})\,
\end{eqnarray}
where \(\gamma = 2s-1\) and the function $F_M(\lambda\,,\varepsilon|\eta)$ is given by
the hypergeometric series
\begin{eqnarray}\label{F}
F_{M}(\lambda\,,\varepsilon|\eta) = \sum_{k=0}^{\infty}
\frac{\Gamma^M(\lambda+k-\varepsilon)}
{\Gamma^M(1+k-\varepsilon)}\,\eta^{k}=\frac{\Gamma (\lambda-\epsilon )^M}{\Gamma (1-\epsilon )^M} \,\times\, _{M+1}F_M(1,\underset{M}{\underbrace{\lambda-\epsilon ,\dots ,\lambda-\epsilon}}  ;\underset{M}{\underbrace{1-\epsilon ,\dots ,1-\epsilon}} ;\eta) \notag
\end{eqnarray}
Therefore we can write in a more compact notation, for \(|\eta|<1\):
\begin{align}\label{Ladder}
I_M =& \frac{2\pi^{M+1} a^M(1-\gamma)}
{(M-1)!\,[\eta]^{\frac{\gamma-1}{2}}}\, \left.\partial_{\varepsilon}^{M-1}\right|_{\varepsilon=0}
\frac{a^M(\gamma+\varepsilon)\Gamma^M(1+\varepsilon)}
{\Gamma^M(1-\varepsilon)}\,
[\eta]^{-\varepsilon}\,
\mathcal{F}_M^{\gamma,\bar{\gamma}}(\eta,\bar \eta|\varepsilon),\nonumber\\
&\text{where}\notag\\
\mathcal{F}_M^{\gamma,\bar{\gamma}}(\eta,\bar \eta|\varepsilon)\,=&\,  {}_{M+1}F_M \left(\left.\begin{matrix}1-\gamma-\varepsilon &  \cdots & 1-\gamma-\varepsilon & \; 1 \\ \;\; 1-\varepsilon & \cdots & 1-\varepsilon &\end{matrix}\; \right| \eta \right)\, {}_{M+1}F_M \left(\left.\begin{matrix}1-\bar \gamma-\varepsilon &  \cdots & 1-\bar \gamma-\varepsilon & \; 1 \\ \;\; 1-\varepsilon & \cdots & 1-\varepsilon &\end{matrix}\; \right| \bar \eta \right).
\end{align}
In the opposite case of \(|\eta|>1\) the same kind of computation can be repeated picking residues in the lower half plane. After redefinition \(n\to -n+ 2 n_s + 2\), this is equivalent to pick the series of poles \(\nu = 2 i s +\frac{in}{2} -ik\,, k=0\,,1\,,2\,,\ldots\), and the residues are
\begin{eqnarray}\label{reslow}
\mathrm{Res}_{\nu = 2 i s +\frac{in}{2} -ik} = \frac{i}{(M-1)!}\eta^{2s} \bar \eta^{2\bar s -2}
\left.\partial_{\varepsilon}^{M-1}\right|_{\varepsilon=0}
\frac{\Gamma^M(1+\varepsilon)\Gamma^M(1-\varepsilon)}
{\Gamma^M(2s+\varepsilon)\Gamma^M(1-2s-\varepsilon)}\,
[\eta]^{\varepsilon}
\\
\nonumber
\frac{\Gamma^M(1-2s+n+k-\varepsilon)}
{\Gamma^M(n+k-\varepsilon)}
\frac{\Gamma^M(2-2\bar{s}+k-\varepsilon)}
{\Gamma^M(1+k-\varepsilon)} \,\eta^{-n-k}\,\bar{\eta}^{-k}\,.
\end{eqnarray}
It follows from \eqref{reslow} that the final expression of the ladder for \(|\eta|>1\) is the same as \eqref{Ladder} after replacing \(\eta\) with \(1/\eta\). For a generic cross-ratio \(|\eta|\lessgtr 1\) the  \(M\)-ladder is, respectively
\begin{align}\label{Ladder_full}
I_M = & \frac{2\pi^{M+1} a^M(1-\gamma)}
{(M-1)!\,[\eta]^{\pm(\frac{\gamma-1}{2})}}\, \left.\partial_{\varepsilon}^{M-1}\right|_{\varepsilon=0}
\frac{a^M(\gamma+\varepsilon)\Gamma^M(1+\varepsilon)}
{\Gamma^M(1-\varepsilon)}\,
[\eta]^{\mp \varepsilon}\,
\mathcal{F}_M^{\gamma,\bar{\gamma}}(\eta^{\pm 1},\bar \eta^{\pm 1}|\varepsilon).
\end{align}
and it shows explicitly the invariance under exchange \(z_1\,\leftrightarrow\, w_1 \); in fact
\begin{align}\label{Ladder_dual}
I_M(\eta)\,=\,I_M\left(\frac{1}{\eta} \right)
\end{align}

The result \eqref{Ladder_full}, obtained under the assumption of \((s,\bar{s})\) in the principal series of \(SL(2,\mathbb{C})\), can be remarkably extended by analytic continuation to \(s=\bar s \in \left(1/2\,,1\right)\), that is setting \(\gamma=\bar \gamma\, \in\, (0,1)\) in \eqref{Ladder_full}. The direct computation of ladder integrals is more involved in this last case, since analytic continuation leads to the failure of the  cancelation of poles by zeros presented on Fig.\ref{poles}, and integration in \eqref{intladder} must be carried out under an appropriate contour deformation prescription. The explicit result for the particular choice of weights \(\gamma=\bar \gamma=1/2 \), corresponding to the isotropic fishnet theory (the case considered by Basso and Dixon in \cite{Basso:2017jwq} for \(D=4\)) reads:
\begin{align}\label{Ladder_iso}
I_M =&\frac{2\pi^{M+1}}
{(M-1)!\,|\eta|^{\pm \frac{1}{2}}}\, \left.\partial_{\varepsilon}^{M-1}\right|_{\varepsilon=0}
\frac{a^M\left(\frac{1}{2}+\varepsilon \right)\Gamma^M(1+\varepsilon)}
{\Gamma^M(1-\varepsilon)}\,
[\eta]^{\mp \varepsilon}\,
\mathcal{F}_M^{\frac{1}{2},\frac{1}{2}}(\eta^{\pm 1},\bar \eta ^{\pm 1}|\varepsilon),\\\nonumber\\
\mathcal{F}_M^{\frac{1}{2},\frac{1}{2}}(\eta,\bar \eta|\varepsilon)\,=&\, _{M+1}F_M \left(\left.\begin{matrix}\frac{1}{2}-\varepsilon &  \cdots & \frac{1}{2}-\varepsilon & \; 1 \\ \;\; 1-\varepsilon & \cdots & 1-\varepsilon &\end{matrix}\; \right| \eta \right)\, {}_{M+1}F_M \left(\left.\begin{matrix}\frac{1}{2}-\varepsilon &  \cdots & \frac{1}{2}-\varepsilon & \; 1 \\ \;\; 1-\varepsilon & \cdots & 1-\varepsilon &\end{matrix}\; \right| \bar \eta \right).\nonumber
\end{align}
Moreover in the isotropic case \(\gamma = 1-\gamma\), and for the simple ``cross" \(N=1\), \(L=1\) diagram (computed below in terms of elliptic functions), the duality \eqref{duality_intro} is a mere consequence of \eqref{Ladder_dual}
\begin{align*}
B^{(1/2)}_{1,1}\left(\eta\right) = I^{(1/2)}_2 (\eta)=I^{(1/2)}_2 \left(\frac{1}{\eta}\right)= B^{(1/2)}_{1,1}\left(\frac{1}{\eta}\right)
\end{align*}
For the sake of duality in the more involved anisotropic case we will need also the relation between ladders with exchange of \(\gamma \,\leftrightarrow\, 1-\gamma \). This relation can be easily checked and looks as follows
\begin{align*}
 I^{\left(1-\gamma \right)}_2 \left(\frac{1}{\eta}\right)\,= \, [\eta]^{\gamma-\frac{1}{2}}[1-\eta]^{1-2 \gamma} \,I_{2}^{\left(\gamma \right)}(\eta),
\end{align*}
and due to \(B^{(\gamma)}_{1,1}\,=\,I_2^{(\gamma)}\) the duality \eqref{duality_intro} is also proved.\\
In the simplest particular case \(M=1\)
 we can simply put $\varepsilon=0$ everywhere and
then reduce to the simple power
\begin{eqnarray}\nonumber
F_{1}(\lambda\,,0|\eta) = \sum_{k=0}^{\infty}
\frac{\Gamma(\lambda+k)}
{k!}\,\eta^{k} = \frac{\Gamma(\lambda)}{(1-\eta)^{\lambda}},
\end{eqnarray}
so that
\begin{eqnarray}\nonumber
G_{L=0,N=1}(z_1\,,w_1\,,z_0) &=&
\,(2\pi^2)^{-1}\left([z_0-z_1]
\,[z_0-w_1]\right)^{\frac{\gamma-1}{2}}B^{(\gamma,\bar \gamma)}_{0,1}(\eta)=\notag \\
\,&=&\left([z_0-z_1]
\,[z_0-w_1]\right)^{\frac{\gamma-1}{2}} \frac{a(1-\gamma,\gamma)}
{[\eta]^{\frac{\gamma-1}{2}}[1-\eta]^{1-\gamma}}
\,=\,\frac{\,1}
{[w_1-z_1]^{1-\gamma}}
\end{eqnarray} which is precisely the single propagator in the trivial case of the Basso-Dixon type formula, with no integrations.
\\
In order to get a better feeling of the structure of our result \eqref{Ladder} at generic \(N+L\), it is instructive to compute the first non-trivial graph \(G_{L=1,N=1}(z_1,w_1,z_0)\) - the two-dimensional ``cross" integral. In four dimensions, the  cross integral  can be computed in terms of the Bloch-Wigner function (di-logarithm function)~\cite{Usyukina:1993ch}. We will see that in our two-dimensional case  the answer for cross can  be expressed through elliptic functions.   Since it involves only \(N=1\) separated variable, it is simply related to the ladder \(I_2\):
\begin{align} \label{crossint}
G_{L=1,N=1}(z_1,w_1,z_0)\,=\,(2\pi^2)^{-1}([z_0-z_1][z_0-w_1])^{\frac{\gamma-1}{2}} I_2(\eta).
\end{align}
For \(M=2\) the ladder integral \eqref{Ladder} reads:
\begin{align*}
&\frac{2\pi^{3} a^2(1-\gamma)}
{[\eta]^{\frac{\gamma-1}{2}}}\,\times\\
\nonumber
&\times \left.\partial_{\varepsilon}\right|_{\varepsilon=0} a^2(\gamma+\varepsilon)\,\frac{\Gamma^2(1+\varepsilon)}
{\Gamma^2(1-\varepsilon)}\,
[\eta]^{-\varepsilon}\,
_{3}F_2 \left(\left.\begin{matrix} 1-\gamma-\varepsilon & & 1-\gamma-\varepsilon & \; 1 \\ \;\; 1-\varepsilon & & 1-\varepsilon &\end{matrix}\; \right| \eta \right)\, {}_{3}F_2 \left(\left.\begin{matrix} 1-\gamma-\varepsilon & & 1-\gamma-\varepsilon & \; 1 \\ \;\; 1-\varepsilon & & 1-\varepsilon &\end{matrix}\; \right| \bar \eta \right)
\end{align*}
Choosing the conformal weights for isotropic fishnets \(\gamma=\bar \gamma =1/2\), the ladder simplifies to
\begin{align}
\label{crossladder}
{2\pi^{3}}\left.\partial_{\varepsilon}\right|_{\varepsilon=0}\,\frac{\Gamma^2(1+\varepsilon)\Gamma^2(1/2-\varepsilon)}
{\Gamma^2(1-\varepsilon)\Gamma^2(1/2+\varepsilon)}\,
[\eta]^{\frac{1}{4}-\varepsilon}\,
_2F_1 \left(\left. \frac{1}{2}-\varepsilon,\frac{1}{2}-\varepsilon;1-2\varepsilon\, \right| \eta \right)\,_2F_1\left(\left. \frac{1}{2}-\varepsilon,\frac{1}{2}-\varepsilon;1-2\varepsilon\,\right| \bar \eta \right)\, .
\end{align}
We can recall the expression of the \(2\)D conformal cross integral \cite{Dotsenko:1984} (e.g. see the formula (1.7) of \cite{Korchemsky1997}); after amputation of one line by sending \(w_0\) to infinity, we get
\begin{align}
\label{crosstilde}
&\tilde G_{h,\bar h}\,=\,\int \frac{ d^2\rho\,}{[w_1-\rho]^h[z_0-\rho]^h [z_1-\rho]^{1-h}}\,=\,\frac{_2F_1(h,h;2h|\eta)\,  _2F_1 (\bar h,\bar h; 2\bar h|\bar \eta)\,[\eta]^h}{[w_1-z_0]^{h}\,B(1-h)}\,+\,(h \leftrightarrow 1-h);\\ \nonumber
&B(h)=\frac{2^{-2i\sigma}(-2 i \sigma)}{\pi} \frac{\Gamma\left(\frac{1}{2}+i\sigma\right)\Gamma\left(-i\sigma\right)}{\Gamma\left(\frac{1}{2}-i\sigma\right)\Gamma\left(i\sigma\right)};\quad \quad h=\frac{1}{2}+i\sigma.
\end{align}
In order to compare with \eqref{crossint} we should set \(h=1/2\), that is \(\sigma=0\). Due to the vanishing of \(B(1/2)\), this expression is an ill-defined sum of two divergent terms. The issue is solved by taking the limit \(\sigma\,\rightarrow\,0\) in \eqref{crosstilde}, which gives the well defined function
\begin{align*}
&  \frac{\pi}{2\, |w_1-z_0|} \,\lim_{\sigma \rightarrow 0}\left[\frac{\Gamma\left(\frac{1}{2}+i\sigma\right)^2\Gamma\left(1-i\sigma\right)^2}{\, \Gamma\left(\frac{1}{2}-i\sigma\right)^2\Gamma\left(1+i\sigma\right)^2}\,[\eta]^{i\sigma} \, F \left(\sigma |\eta \right)\,F \left(\sigma | \bar \eta \right)\, \,+\,(\sigma \leftrightarrow -\sigma) \right]\, ,\\
&\text{where}\;\;F \left(\sigma | x \right)\,=\, _2F_1\left(\left. \frac{1}{2}+i \sigma,\frac{1}{2}+i\sigma ;1+ 2i \sigma\, \right| x \right)\,
\end{align*}
and reproduces the result of plugging  \eqref{crossladder} into \eqref{crossint}. The problem reduces to computing \(F \left(\sigma |\eta \right)\) and \(\partial_\sigma|_{\sigma=0}F\left(\sigma |\eta \right)\) which  reduce to elliptic integrals. Then the cross integral can be presented in explicit form:
\begin{align}
&I_{1,1}^{{BD}}(z_0,z_1,w_0,w_1)\, \equiv \, \int \frac{ d^2\rho\,}{|z_0-\rho| |w_0-\rho| |z_1-\rho| |w_1-\rho|}\,=\,\notag\\
&=\frac{4  \,|1-\eta|\,}{ |w_1-z_1||w_0-z_0|} \left[ K(\eta)K(1-\bar \eta)+   \,K(\bar \eta )K(1-\eta)\right], \quad \; |\eta|<1
\end{align}
where here:
\begin{align*}
\eta=\frac{z_0-w_1}{w_1-w_0}\frac{z_1-w_0}{z_0-z_1}
\end{align*}
and \(K(x)\) is the elliptic K integral:
\begin{align*}
K(x) = \int_0^1  \frac{dt}{\sqrt{(1-t^2)(1-x\, t^2)}}.
\end{align*}

This result for the cross integral suggests that even for any \(L,N\) the formula for two-dimensional Basso-Dixon integral can be presented in terms elliptic poly-logarithms encountered~\cite{2007math......3237L} in various Feynman graph calculations.    

\section{The case of ladders \(N=1\), \(L>1\) and the simple wheel integral }

The computation of \(2\)-dimensional ladders carried out in the previous sections has other interesting applications in the context of the theory \eqref{bi-scalarL}.  The simplest observables in this theory are single trace operators \(\tr(X^l)(z)\), \(\tr(Z^l)(z)\). As explained in~\cite{Gurdogan:2015csr,Kazakov:2018qez}, the perturbative expansions of their correlators\begin{align}
\langle\tr X^l(z)\tr(X^\dagger)^l(w)\rangle\,\quad\; \langle \tr Z^l(z)\tr(Z^\dagger)^l(w)\rangle \label{correlators}
\end{align}
consist, for \(l>2\), of only of the ``globe"-shaped fishnet Feynman integrals:
\begin{align*}
F_{l,N}(x,y)\,= \,\int \prod^l_{j=1} \frac{1}{|z_{0,j}-z_{j,1}|^{1+2\omega}|z_{j,N}-z_{j,N+1}|^{1+2\omega}}\prod^N_{k=1}\frac{d^2 z_{j,k}}{|z_{j,k}-z_{j,k+1}|^{1+2\omega}|z_{j,k}-z_{j+1,k}|^{1-2\omega}},
\end{align*}
where we set \(z_{j,0}\equiv z,\;z_{j,N+1} \equiv w\), and the expansion itself reads:
\begin{equation}
G_l(z-w)\,=\,\sum_{N=0}^{\infty}\,\xi^{2Nl}\,F_{l,N}(z,w)
\end{equation}
For any value of the coupling \(\xi^2\) the correlators \eqref{correlators} are conformal, thus it is possible to define the scaling dimension of the fields \(X\) and \(Z\) as:
\begin{align}
\Delta(\xi^2)\,=\,\,- \lim_{|z-w|\to \infty}\frac{\log(G_l(z-w))}{\log(z-w)^2}\,=\,\frac{l}{2} + \gamma(\xi^2)
\end{align}
where the anomalous dimension \(\gamma\) is an expansion in the log-divergence of \(F_{l,N}\) graphs, i.e. its coefficient of \(\frac{1}{\varepsilon}\) in dimensional regularization. 
\begin{figure}[t]
\centerline{\includegraphics[scale=1.4]{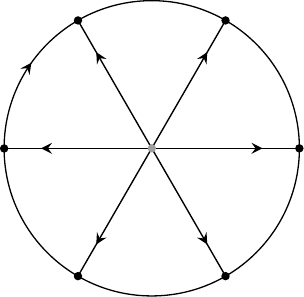}}
\caption{Simple wheel at \(l=6\). The black blobs are integrated over, while the gray blob in the center of the figure is the external point of \(F_{l,N}(z,w)\) left over after amputation.}
\label{simpwheel}
\end{figure}
Since this divergence is the same for the corresponding wheel graph, obtained after amputation of \(|z_{j,N}-z_{j,N+1}|\) propagators:
\begin{align}
W_{l,N}(z)\,= \,\int \prod^l_{j=1} \frac{1}{|z_{0,j}-z_{j,1}|^{1+2\omega}}\prod^N_{k=1} \frac{d^2 z_{j,k} }{|z_{j,k}-z_{j,k+1}|^{1+2\omega}|z_{j,k}-z_{j+1,k}|^{1-2\omega}}, \label{wheel}
\end{align}
we can write
\begin{align*}
-\gamma(\xi^2)\,=\,\sum_{N=1}^{\infty}\xi^{2Nl}\, W^{(1)}_{l,N}
\end{align*}
where \(W_{l,N}^{(1)}\) stands for the \(1/\varepsilon\)-divergence coefficient in the expansion of the \((l,N)\) wheel in dimensional regularization.\footnote{In general, the following wheel integral has \(\frac{1}{\epsilon^N}\) divergency, so one has to extract the subleading \(\frac{1}{\epsilon}\) term.}
~The simple case \(N=1\) can be worked out explicitly, since the integral \eqref{wheel} can be regarded as a ladder with periodic boundary conditions and \(L=l-1\), see Fig.\ref{simpwheel}. In the formalism of integral operators \eqref{gen_comb} we can write:
\begin{align}
W_{l,1}(z)\,= \,\int \prod^l_{j=1} \frac{d^2 z_{j} }{[z_0-z_j]^{2s-1}[z_{j}-z_{j+1}]^{2-2s}}\,=\,\text{Tr}\left[\Lambda^l_1(x|z_0)\right],
\label{trace_wheel}
\end{align}
where \(x=s-1,\,s=\bar s =3/2-\omega\). We can insert inside the trace in \eqref{trace_wheel} a complete basis \eqref{SCS} in order to get an integral over one separated variable:
\begin{align}
\frac{1}{2\pi^2}\,\sum_{n=-\infty}^{\infty}\int_{-\infty}^{+\infty} d\nu\, \text{Tr}\left[\Lambda^l_1(x|z_0) \Psi(x|z)\overline{\Psi(x|z')}\right]\,=\,\frac{1}{2\pi^2}\,\left(\sum_{n=-\infty}^{\infty}\int_{-\infty}^{+\infty} d\nu\,\lambda^l_1(x)\right)\,\int d^2 z \, \Psi(x|z)\overline{\Psi(x|z)}.
\end{align}
The integration over \(z\) is the scalar product of two eigenfunctions with the same weights \(x\), thus carrying the \(\log\)-divergence of \eqref{trace_wheel}, or the  \(\frac{1}{\epsilon}\) divergence which is the leading one at \(N=1\) in the \(\epsilon\)-regularization.  We can easily extract it:
\begin{align*}
\int_{UV} d^{2+\epsilon} z \, \Psi(x|z)\overline{\Psi(x|z)}\,=\,2\pi \int_{0}^{1} \frac{dr}{r^{1-\epsilon}}\,=\,\frac{2\pi}{\varepsilon}\, ,
\end{align*}
and the resulting \(W^{(1)}_{l,1}\) reads:
\begin{align*}
W^{(1)}_{l,1}\,=\, \frac{1}{2\pi^2}\sum_{n=-\infty}^{\infty}\int_{-\infty}^{+\infty} d\nu\,\lambda^l_1(x) \,=\,\frac{1}{\pi}\, I_{l}(\eta)|_{ \eta=1}
\end{align*}
The \(L\)-ladder at \(\eta=1\) is a finite quantity only for \(L=l-1>1\), and it isn't otherwise possible to close the integration contour in \eqref{intladder}. Indeed the asymptotic expansion of \(\lambda_1\) in \(\nu\) is
\begin{align*}
\lambda^{L+1}_1(n,\nu)\,=\, (-i \nu)^{-L-1} + O(\nu^{-L})\, .
\end{align*}
The divergence of the wheel diagram at \(l=L+1=2\) is in agreement with our expectations: in order to renormalize correlators \eqref{correlators} at \(l=2\) the specific double-trace counterterms are needed  \cite{Fokken:2014soa,Fokken:2013aea,Sieg:2016vap,Kazakov:2018qez,Grabner:2017pgm,Gromov:2018hut}.\\
More explicitly, fixing the propagators along the frames and spokes to be the same (\(\omega=0\)), we get:
\begin{align}
\label{anomal_dim}
&W^{(1)}_{l,1}\,=\,\frac{2\pi^{l}}
{(l-1)!}\, \left.\frac{d^{l-1}}{d{\varepsilon}^{l-1}}\right|_{\varepsilon=0}\,
\frac{\Gamma^{l}(1+\varepsilon)\Gamma^{l}(1-\varepsilon)}
{\Gamma^{l}(3/2+\varepsilon)\Gamma^{l}(-1/2-\varepsilon)}\,
\left(\,\sum_{k=0}^{\infty}
\frac{\Gamma^{l}(1/2+k-\varepsilon)}
{\Gamma^{l}(1+k-\varepsilon)}\right)^2\,
\end{align}

The quantity \eqref{anomal_dim} can be computed numerically and, hopefully, expressed in terms of Elliptic Multiple Zeta Values.

\section{Conclusions and prospects }
In this paper, we derived an explicit formula for the two-dimensional analogue of Basso-Dixon   integral given by  conformal  fishnet  Feynman graph represented by regular square lattice of rectangular  \(L\times N\) shape, presented on Fig.\ref{BDgraph} and Fig.\ref{BDtransf}(left).  The definition of this integral and the  result are presented at the end of Introduction~(sec.\ref{intro}). Our result  represents  a slightly more general quantity then Basso-Dixon graph: it concerns the anisotropic fishnet, i.e. with different powers for vertical and horizontal propagators, corresponding to arbitrary spins \(s,\bar s\) of principal series representation of \(SL(2,\mathbb{C})\) group, or for the analytic continuation to \(s=\bar s\) belonging to the real interval \(\left(\frac{1}{2},1\right)\). The particular case of  isotropic fishnet, a-la Basso-Dixon, corresponds to the case \(s=\bar s=3/4\). In two-dimensional case the fishnet graph is built from propagators \(\frac{1}{|z_1-z_2|}\).  Such graph is a particular case of  single-trace correlators introduced in~\cite{Chicherin:2017cns,Chicherin:2017frs} for the study of planar scalar scattering amplitudes in the bi-scalar fishnet CFT  \cite{Gurdogan:2015csr,Kazakov:2018qez}. In the simplest case \(N=L=1\) (cross integral) we managed to present the result in terms of elliptic functions. It seems plausible that even for general \(L,N\) the result can be expressed in terms of elliptic functions. A probable full basis of such functions, in terms of which our quantity could be presented, are the so-called multiple elliptic poly-logarithmic functions (see \cite{Passarino:2016zcd} and references therein). It would be interesting to obtain it for a few smallest \(N,L\).  

Interestingly, in the case \(s\to 1/2\) (or, alternatively, \(s\to 1\), which is an equivalent \(SL(2,\mathbb{C})\) representation for the graph's propagators) this fishnet corresponds to one of the conservation laws of Lipatov integrable (open) spin chain hamiltonian \cite{Bartels:2011nz, Lipatov:2009nt} describing the system of reggeized gluons for the Regge (BFKL) limit of QCD ~\cite{,Lipa:1993pmr,Faddeev:1994zg,Derkachov:2001yn, DeVega:2001pu}. It would be interesting to understand what kind of BFKL physics it can describe. 

The Basso-Dixon type configuration represents only one set of possible physical quantities which can be, in principle, analyzed and computed in the planar bi-scalar fishnet CFT due to integrability. To fix the OPE rules in such a theory, we have to compute the spectrum of anomalous dimensions and the structure constants of all local operators. Some of them have been analyzed and even computed in the literature. In particular, the so-called wheel graphs, corresponding to operators \(\tr X^L\), have been computed  in \(D=4\) dimensions in \cite{Ahn:2011xq,Gurdogan:2015csr} up to two wrappings at any \(L\).  In \cite{Gromov:2017cja}    they have been computed in particular cases of \(L=2,3\) (\(L=4\) case is to appear \cite{GrabnerGromovKazakovKorchemsky}) to any reasonable loop order (for any wrapping there exists an iterative analytic procedure) or numerically with a great precision,  by means of the Quantum Spectral Curve method \cite{Gromov:2013pga,Gromov:2014caa,Gromov:2017blm,Kazakov:2018ugh}. We think that, to give a more general result  for any \(L\) in rather explicit form, we have to employ a  powerful technique of separated variables, similarly to the one we employed here in two dimensions for Basso-Dixon type graphs. The first step would be to compute the wheel graphs in two dimensions using the techniques of this paper. To advance to \textbf{ \(D>2\) }  dimensions by integrable spin chain methods,  we have to understand the construction of separated variables for higher rank symmetries, such as \(SU(2,2)\). Some recent results in this direction might provide the necessary computational tools~\cite{Gromov:2016itr, Derkachov:2018ewi, Ryan:2018fyo, Maillet:2018czd, Maillet:2018rto, Maillet2018}. It would be also good to generalize our techniques, at least in two dimensions, to the computation  of multi-magnon operators related to ``multi-spiral" Feynman graphs~\cite{Caetano:2016ydc}.

The computation of structure constants is an even more complicated task. Certain explicit  results for OPE of short protected operators  have been obtained for fishnet CFT in \cite{Grabner:2017pgm,Gromov:2018hut,Kazakov:2018qez} (see also \cite{Balitsky:2015oux,Balitsky:2015tca} in BFKL limit) using solely the conformal symmetry. The calculation of more complicated structure constant is  a  difficult task demanding the most sophisticated techniques, such as SoV method.   Since for the \(2D\) case the SoV formalism is well developed it would be interesting to apply the methods of the current paper to computations of more complicated  structure constants at least in two dimensions.

Finally, it would be good to understand the role of separated variables in the non-perturbative structure of the bi-scalar fishnet CFT. A good beginning would be to understand in terms of SoV the strong coupling limit for long operators of the theory and to relate it to the classical limit of the dual non-compact sigma model which will probably arise in two-dimensional case similarly to the one which was observed in  four-dimensional bi-scalar fishnet CFT in~\cite{Basso:2018agi}.

\section*{Acknowledgements}

We are thankful to B. Basso, J. Caetano, F.~Levkovich-Maslyuk, D.~Zhong, G.~Ferrando ~for
discussions. Our work  was supported   by the European Research Council (Programme
``Ideas" ERC-2012-AdG 320769 AdS-CFT-solvable). The work of S.D. is supported by the Russian Science Foundation (project no.14-11-00598). The work of E.O. is supported by the German Science Foundation (DFG) under the Collaborative Research Center (SFB) 676 Particles, Strings and the Early Universe and the Research Training Group 1670.

\section*{\Large Appendices}

\appendix

\section{Diagram technique}\label{app_A1}

The functions and kernels of integral operators
considered in the main body of the paper are represented in the form of
two-dimensional Feynman diagrams. The propagator which is shown by the arrow directed from $w$ to $z$ and
index $\alpha$ attached to it is given by the following expression
\begin{equation}
\frac{1}{[z-w]^\alpha}\equiv\frac{1}{(z-w)^\alpha (z^*-w^*)^{\bar\alpha}}=
\frac{(z^*-w^*)^{\alpha-\bar\alpha}}{|z-w|^{2\alpha}}\,,
\end{equation}
where the difference $\alpha-\bar\alpha$ is integer:
$\alpha-\bar\alpha \in \mathbb{Z}$.\footnote{Note that the star \(^*\) is used for the usual complex conjugation whether as the meaning of the bar is explained in eq.\eqref{bar_s},\eqref{bar_x}.}
The flip of the arrow in propagator gives an additional sign factor
$(-1)^{\alpha-\bar \alpha}$ for which we shall use the shorthand notation
\begin{eqnarray}
\label{signnotation}
(-1)^{[\alpha]} = (-1)^{\alpha-\bar \alpha}
\end{eqnarray}
so that
\begin{equation}
\frac{1}{[z-w]^\alpha}=
\frac{(-1)^{\alpha-\bar\alpha}}{[w-z]^{\alpha}}= \frac{(-1)^{[\alpha]}}{[w-z]^{\alpha}}\,.
\end{equation}
The evaluation of  Feynman diagrams is based on
their transformation with the help of the certain rules, namely:
\begin{itemize}
\item Chain  relation:
\begin{eqnarray}\label{Chain}
\int d^2 w\frac{1}{[z_1-w]^\alpha [w-z_2]^{\beta}}=
(-1)^{[\gamma]}a(\alpha,\beta,\gamma)
\frac{1}{[z_1-z_2]^{\alpha+\beta-1}}\,,
\end{eqnarray}
where $\gamma=2-\alpha-\beta,\ \bar\gamma=2-\bar\alpha-\bar\beta$.
\item Special case of the chain relation
\begin{eqnarray}\label{Delta2}
\int d^2 w\frac{1}{[z_1-w]^{1-\alpha}[w-z_2]^{1+\alpha}}=-\pi^2\, \frac{(-1)^{[\alpha]}}
{\alpha\bar\alpha}\, \delta^2(z_1-z_2)\,,
\end{eqnarray}

\item Star--\,triangle relation:
\begin{eqnarray}\label{Star}
\int d^2w\frac{1}{[z_1-w]^\alpha[z_2-w]^\beta [z_3-w]^\gamma}=
\frac{\pi a(\alpha,\beta,\gamma)}{[z_2-z_1]^{1-\gamma}[z_1-z_3]^{1-\beta}[z_3-z_2]^{1-\alpha}}\,,
\end{eqnarray}
where $\alpha+\beta+\gamma=2$ and $\bar\alpha+\bar\beta+\bar\gamma=2$.
\item Cross relation:
\begin{eqnarray}\label{Cross}
\frac{1}{[z_1-z_2]^{\alpha'-\alpha}}\int d^2w
\frac{a(\alpha',\bar\beta')}{[w-z_1]^\alpha[w-z_2]^{1-\alpha'}
[w-z_3]^\beta [w-z_4]^{1-\beta'}}=
\nonumber\\
=\frac{1}{[z_3-z_4]^{\beta'-\beta}}
\int d^2\zeta
\frac{a(\alpha,\bar\beta)}{[w-z_1]^{\alpha'}[w-z_2]^{1-\alpha}
[w-z_3]^{\beta'} [w-z_4]^{1-\beta}}\,,
\end{eqnarray}
where $\alpha+\beta=\alpha'+\beta'$.
\end{itemize}

\begin{figure}[t]
\centerline{\includegraphics[width=0.7\linewidth]{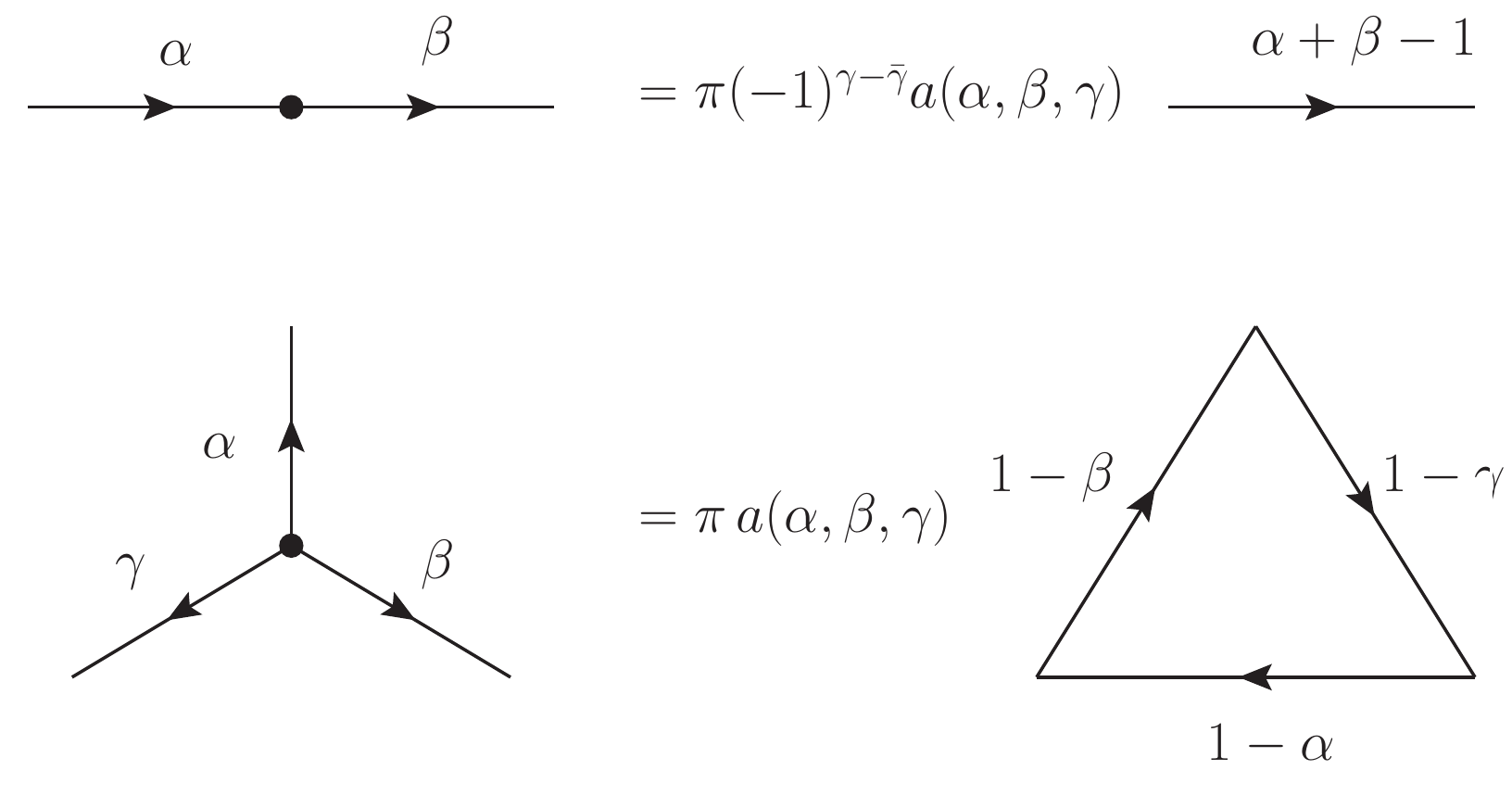}}
\caption{The chain and star--\,triangle relations, $\alpha+\beta+\gamma=2$.}
\label{Rules}
\end{figure}

\begin{figure}[t]
\centerline{\includegraphics[width=0.9\linewidth]{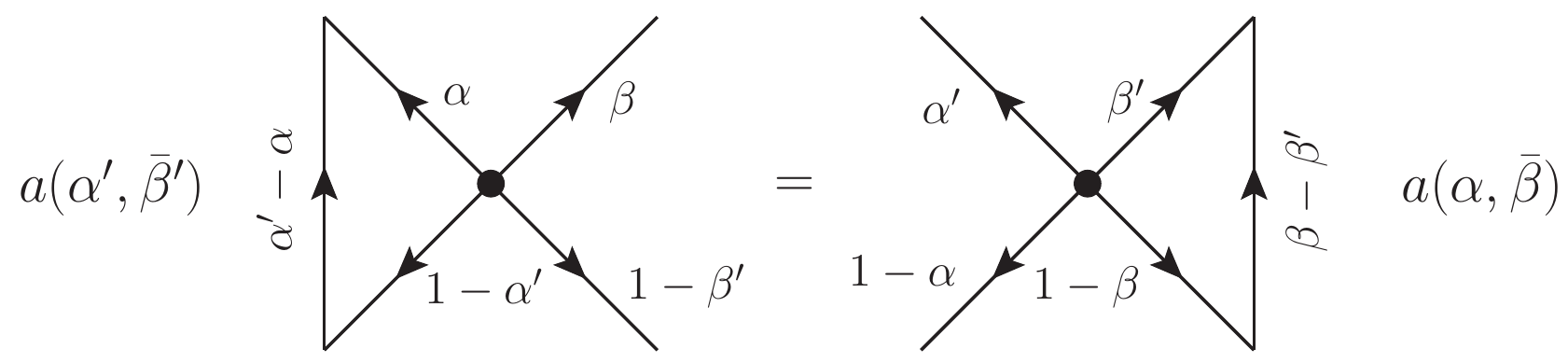}}
\caption{The cross relation, $\alpha+\beta=\alpha'+\beta'$.}
\label{fig:Cross}
\end{figure}

These relations are shown in
diagrammatic form in Figs.~\ref{Rules}, \ref{fig:Cross}.
Here the notation
$a(\alpha,\beta,\gamma,\ldots)=a(\alpha)a(\beta) a(\gamma)\ldots$ is
introduced for the product of special function $a(\alpha)$
for different values of arguments.
The definition of the function $a(\alpha)$ is the following
\begin{equation}
a(\alpha)=\frac{\Gamma(1-\bar\alpha)}{\Gamma(\alpha)}.
\end{equation}
Note that this function depends on two parameters $\alpha$ and $\bar{\alpha}$, where the difference $\alpha-\bar\alpha$ should be integer, but for the sake of  simplicity we shall use the shorthand notation $a(\alpha)$.
There are some useful relations for this function
\begin{equation}
a(1+\alpha) = -\frac{a(\alpha)}{\alpha\bar\alpha}\,, \quad
a(\alpha)a(1-\alpha)=(-1)^{[\alpha]}\,,\quad
a(1+\alpha)a(1-\alpha) = -\frac{(-1)^{[\alpha]}}
{\alpha\bar\alpha}
\end{equation}
\section{Reduction and duality}\label{reductions}

We start from the simplest
example \(N=1\,,L=1\), make the reduction by sending \(w_0\to\infty\) and drop the corresponding propagator. 
We want to reduce the original quantity
\begin{align*} &I^{BD}_{1,1}(z_0,z_1,w_0,w_1)=\int d^2w \frac{1}{[w-z_1]^{1-\gamma}[w_1-w]^{1-\gamma}
[w-w_0]^{\gamma}[z_0-w]^{\gamma}}\to \\
&\to G_{1,1}(z_1,w_1|z_0)=\int d^2w \frac{1}{[w-z_1]^{1-\gamma}[w_1-w]^{1-\gamma}
[z_0-w]^{\gamma}}
\end{align*}   We can always restore the original quantity  \(I^{BD}_{1,1}(z_0,z_1,w_0,w_1)\) from \(G_{1,1}(z_1,w_1|z_0)\) using  its conformal symmetry, i.e. by  applying the shift+inversion transformation: \begin{align*}  
& G_{1,1}\left(\left. \frac{1}{{z}_1},\frac{1}{{w}_1} \right|\frac{1}{{z}_0}\right)= \int \frac{d^2w}{[w]^2} \frac{1}{[1/w-1/z_1]^{1-\gamma}[1/w_1-1/w]^{1-\gamma}
[1/z_0-1/w]^{\gamma}}=\\
&=[z_1]^{1-\gamma}[w_1]^{1-\gamma}[z_0]^{\gamma} \int  \frac{d^2w\,\,}{[w]^{\gamma}[z_1-w]^{1-\gamma}[w-w_1]^{1-\gamma}
[w-z_0]^{\gamma}}\\
&=[z_1]^{1-\gamma}[w_1]^{1-\gamma}[z_0]^{\gamma}\,I^{BD}_{1,1}(z_0,z_1,0,w_1)
\\
&=[z_1]^{1-\gamma}[w_1]^{1-\gamma}[z_0]^{\gamma}\,I^{BD}_{1,1}(z_0+w_0,z_1+w_0,w_0,w_1+w_0)
\end{align*} or    
\begin{align*}
 I^{BD}_{1,1}(z_0,z_1,w_0,w_1)=[z_1-w_0]^{\gamma-1}[w_1-w_0]^{\gamma-1}[z_0-w_0]^{{-\gamma}}G_{1,1}\left(\left.\frac{1}{{z}_1-{w}_0},\frac{1}{{w}_1-{w}_0} \right|\frac{1}{{z}_0-{w}_0}\right).
\end{align*}

Analogously, the formula for the general \(N,L\) looks  as follows:\begin{align}
 I^{BD}_{L,N}(z_0,z_1,w_0,w_1)=[z_1-w_0]^{N(\gamma-1)}[w_1-w_0]^{N(\gamma-1)}[z_0-w_0]^{{-L \gamma}}G_{L,N}\left(\left.\frac{1}{{z}_1-{w}_0},\frac{1}{{w}_1-{w}_0} \right|\frac{1}{{z}_0-{w}_0}\right),
 \label{reducedeq}
\end{align}
where 
\begin{eqnarray}\label{BDint2Dapp}
I_{L,N}^{\text{BD}}(z_0,z_1,w_0,w_1)  \,=\int\,  \prod_{l=1}^L\,\prod_{n=1}^N\, d^2z_{ln}\left( \prod_{(l,n)\in {\cal L}_{L,N}}\, \frac{1}{|z_{l,n}-z_{l,n+1}|^{1+2\omega}
\, | z_{l,n}-z_{l+1,n}|^{1-2\omega}}\right).\qquad 
\end{eqnarray}
Taking into account \eqref{BDSOVgen} and \eqref{BDSOV} and setting:
\begin{align*}
z_1'=(z_1-w_0)^{-1},\quad z_0'=(z_0-w_0)^{-1},\quad w_1'=(w_1-w_0)^{-1},\;\;\text{and}\;\;\eta'= \frac{w_1'-z_0'}{z_1'-z_0'}
\end{align*}
we can give an explicit expression for the last factor in \eqref{reducedeq} in terms of function \(B_{L,N}(\eta)\):
\begin{align*}
G_{L,N}(z_1',w_1'|z_0')\,=&\,([z_0'-z_1'][z_0'-w_1'])^{N\frac{\gamma-1}{2}} B_{L,N}\left(\eta '\right)\\\\=&\, \left(\frac{[z_0-z_1]}{[z_0-w_0][z_1-w_0]}\right)^{(\gamma-1)N} \,[\eta]^{\frac{\gamma-1}{2} N} B_{L,N}\left(\eta \right)
\end{align*}
where \(\eta\) is the anharmonic ratio of the graph \(I_{L,N}^{BD}\):
\begin{equation}
\eta=\frac{(w_1-z_0)(z_1-w_0)}{(z_1-z_0)(w_1-w_0)}
\end{equation}
By definition \eqref{BDint} our graphs should have a duality symmetry, namely:
\begin{align}
\label{duality}
I^{BD}_{L,N}(z_0,z_1,w_0,w_1)\,=\, I^{BD}_{N,L}(z_1,w_0,w_1,z_0)
\end{align} 
Namely we can rotate the whole diagram anti-clockwise by an angle $\frac{\pi}{2}$ and repeat our computation by eigenfunction expansion step by step with some changes:
\begin{itemize}
\item $L \rightleftarrows N $
\item $\gamma \rightleftarrows 1-\gamma$, so that now
horizontal lines have index $\gamma$ and vertical $1-\gamma$
\end{itemize}
and we derive a different representation for the same quantity
\begin{align}
\label{BDdual}
& I^{BD}_{N,L}(z_0, z_1,w_0,w_1) = \,\frac{[w_0-z_1]^{-\gamma L}[z_0-w_1]^{-\gamma L}}
{[z_1-w_1]^{-\gamma L+(1-\gamma) N}}\,
\left[{\eta}\right]^{\frac{\gamma}{2}L}\, B^{(1-\gamma)}_{N,L}
\left(\frac{1}{\eta}\right)
\end{align}
\section{Details of the derivation of the formula~(\ref{res})}\label{app_B}

The derivation of the formula~(\ref{res}) contains three steps:
\begin{itemize}
\item calculate integrand at $\nu = \frac{in}{2} +ik-i\varepsilon$
\begin{eqnarray}\label{k}
(-1)^{Mn}\frac{\Gamma^M(2-2\bar{s}+k-\varepsilon)\Gamma^M(-k+\varepsilon)}
{\Gamma^M(2s-n-k+\varepsilon)\Gamma^M(n+k-\varepsilon)} \,\eta^{n+k-\varepsilon}\bar{\eta}^{k-\varepsilon}
\end{eqnarray}
\item use twice the Euler reflection formula
\begin{eqnarray}\nonumber
\Gamma(-k+\varepsilon) = \frac{1}{\varepsilon}
\frac{(-1)^k\,\Gamma(1+\varepsilon)\Gamma(1-\varepsilon)}
{\Gamma(1+k-\varepsilon)}\,,
\\
\nonumber
\frac{1}{\Gamma(2s-n-k+\varepsilon)} =
\frac{(-1)^{n+k}\Gamma(2s+\varepsilon)\Gamma(1-2s-\varepsilon)}
{\Gamma(1-2s+n+k-\varepsilon)}\,,
\end{eqnarray}
to transform~(\ref{k}) to the form
\begin{eqnarray}\nonumber
\frac{1}{\varepsilon^M}
\frac{\Gamma^M(1+\varepsilon)\Gamma^M(1-\varepsilon)}
{\Gamma^M(2s+\varepsilon)\Gamma^M(1-2s-\varepsilon)}
\frac{\Gamma^M(1-2s+n+k-\varepsilon)}
{\Gamma^M(n+k-\varepsilon)}\,
\frac{\Gamma^M(2-2\bar{s}+k-\varepsilon)}
{\Gamma^M(1+k-\varepsilon)}\,
\eta^{n+k-\varepsilon}\bar{\eta}^{k-\varepsilon}
\end{eqnarray}
\item extract the coefficient in front of $\frac{1}{\varepsilon}$
and multiply it by $(-i)$.
\end{itemize}




\bibliographystyle{JHEP}
\bibliography{biblio_v2}

\end{document}